\documentclass[amsmath,amssymb,aps,prd,11pt,superscriptaddress,nofootinbib,preprint]{revtex4-1}
\pdfoutput=1
\usepackage{slashed}
\usepackage{graphicx}
\usepackage{subfigure}
\usepackage{soul}

\usepackage{amsmath}
\usepackage{multirow}
\usepackage{tablefootnote}
\usepackage{hyperref}
\usepackage{epstopdf}
\usepackage{comment}
\usepackage[utf8]{inputenc}

\usepackage[dvipsnames]{xcolor}

\newcommand{\equa}{equation}
\newcommand{\beq}{\begin{\equa}}
\newcommand{\eeq}{\end{\equa}}
\newcommand{\eqna}{eqnarray}
\newcommand{\bea}{\begin{\eqna}}
\newcommand{\eea}{\end{\eqna}}

\def\figureautorefname~#1\null{Fig.\,#1\null}
\def\tableautorefname~#1\null{Table\,#1\null}
\def\sectionautorefname~#1\null{Section\,#1\null}
\def\subsectionautorefname~#1\null{Section\,#1\null}
\def\equationautorefname~#1\null{Eq.\,(#1)\null}

\def\m1{M_1}
\def\m2{M_2}
\def\m3{M_3}

\def\ch10{\tilde \chi^0_1}

\def\gev{\,{\rm GeV}}

\def\to{\rightarrow}

\newcommand{\lsim}{\mathrel{\mathop{\kern 0pt \rlap
  {\raise.2ex\hbox{$<$}}}
  \lower.9ex\hbox{\kern-.190em $\sim$}}}
\newcommand{\gsim}{\mathrel{\mathop{\kern 0pt \rlap
  {\raise.2ex\hbox{$>$}}}
  \lower.9ex\hbox{\kern-.190em $\sim$}}}

\definecolor{pink}{RGB}{255,105,180}

\def\cosba{\cos(\beta-\alpha)}

\newcommand{\mm}{{\mu^{+} \mu^{-}}}
\newcommand{\ee}{{e^{+} e^{-}}}

\newcommand{\tanb}{\tan \beta}

\allowdisplaybreaks[4]

\newcommand{\GeV}{\textrm{ GeV}}

\RequirePackage[normalem]{ulem}

\begin{document}
\preprint{PITT-PACC-2102, ADP-21-2/T1149}

%\hspace{5.2in}\mbox{PITT-PACC-2102}

%\hspace{5.2in}\mbox{ADP-21-2/T1149}
\title{Heavy Higgs Bosons in 2HDM at a Muon Collider}

\author{Tao Han}
\email{than@pitt.edu}
\affiliation{Department of Physics and Astronomy, University of Pittsburgh,  Pittsburgh, PA 15260, USA}
\author{Shuailong Li}
\email{shuailongli@email.arizona.edu}
\affiliation{Department of Physics, University of Arizona, Tucson, Arizona  85721, USA}
\author{Shufang Su}
\email{shufang@email.arizona.edu}
\affiliation{Department of Physics, University of Arizona, Tucson, Arizona  85721, USA}
\author{Wei Su}
\email{wei.su@adelaide.edu.au}
\affiliation{ARC Centre of Excellence for Dark Matter Particle Physics, Department of Physics, University of Adelaide, South Australia 5005, Australia}
\author{Yongcheng Wu}
\email{ywu@okstate.edu}
\affiliation{Ottawa-Carleton Institute for Physics, Carleton University, \\1125 Colonel By Drive, Ottawa, Ontario K1S 5B6, Canada}
\affiliation{Department of Physics, Oklahoma State University, Stillwater, OK, 74078, USA}

\begin{abstract}
    {We study the discovery potential of the  non-Standard Model (SM) heavy Higgs bosons in the Two-Higgs-Doublet Models  (2HDMs) at a multi-TeV muon collider and explore the discrimination power among  different types of 2HDMs.  We find that the pair production of the non-SM Higgs bosons via the universal gauge interactions is the dominant mechanism once above the kinematic threshold.
    %when their masses are below half of the center of mass energies.
    Single Higgs boson production associated with a pair of heavy fermions could be important in the parameter region with enhanced Yukawa couplings. For both signal final states,  $\mm$ annihilation channels dominate over the vector boson fusion (VBF) processes, except at high center of mass energies where the VBF processes receive large logarithmic enhancement with the increase of energies. Single Higgs boson $s$-channel production in $\mm$-annihilation via the radiative return can also be important for the Type-L 2HDM in the very large $\tan\beta$ region, extending the kinematic reach of the heavy Higgs boson mass to the collider energy. Considering both the production and decay of non-SM Higgs bosons, signals can be identified over the Standard Model backgrounds. With the pair production channels via annihilation, 95\% C.L. exclusion reaches in the Higgs mass up to the production mass threshold of $\sqrt{s}/2$ are possible when channels with different final states are combined.  Including single production modes can extended the reach further. Different types of 2HDMs can be distinguishable for moderate and large values of $\tan\beta$.}
\end{abstract}
% \keywords{muon collider, 2HDM.}
% \preprint{PITT-PACC-2007  ADP-20-24/T1134}
\maketitle
% \preprint{
% \begin{flushright}
% PITT-PACC-2102  \\
% ADP-21-2/T1149
% \end{flushright}
% }
\tableofcontents
\flushbottom
\clearpage

%%%%%%%%%%%%%%%%%%%%%%%%%%%%%%%%%%%%%%
\section{Introduction}

The discovery of the Standard Model Higgs boson completes the particle spectrum of the Standard Model (SM) of elementary particle physics. Yet, there are still unsolved mysteries,  including theoretical considerations such as the mechanism to stabilize the electroweak scale and the neutrino mass generation, and observations such as the nature of the particle dark matter and the matter-antimatter asymmetry. Thus, there are strong motivations to consider theories beyond the SM (BSM) near the TeV scale.   Many theories beyond the SM naturally contain an extended Higgs sector. The most common incarnation is the two-Higgs-double model (2HDM). Including one more electroweak (EW) Higgs doublet, the theory leads to rich phenomenology of the new Higgs bosons and flavor physics. The searches for   new Higgs bosons have been actively carrying out at colliders, most notably at the LHC (see Ref.~\cite{Kling:2020hmi} for a summary and references therein).  The absence of the signal observation leads to the current bounds on the mass and couplings of those non-SM Higgs bosons.    Future high-energy $\ee$ colliders such as the 1-TeV International Linear Collider (ILC) and the 3-TeV Compact Linear Collider (CLIC) will be able to extend the mass coverage close to half of the center-of-mass (c.m.) energy for pair production of new particles.
Significant improvements have been anticipated at the future high-energy hadron collider such as the 100 TeV $pp$ Future Circular Collider (FCC) and the Super Proton Proton Collider (SPPC), reaching a mass coverage of $2-4$ TeV for the heavy Higgs bosons via the exotic decay modes~\cite{Kling:2018xud,Li:2020hao} and $5-20$ TeV via conventional decay modes of $tt$, $bb$, $tb$, and $\tau\tau$~\cite{Hajer:2015gka,Craig:2016ygr}.    Indirect limits on the heavy Higgs masses can also be obtained via the precision measurements of the SM-like Higgs couplings at future Higgs factories, when loop corrections are included~\cite{Gu:2017ckc, Chen:2018shg, Chen:2019pkq, Han:2020lta}.

Recently, there have been renewed interests for muon colliders operating at high energies in the range of multi-TeV  \cite{Delahaye:2019omf,Han:2020uid,Long:2020wfp}. This would offer great physics opportunity to open  unprecedented new energy threshold for new physics, and provide precision measurements in a clean environment in leptonic collisions~\cite{Capdevilla:2021rwo,Liu:2021jyc,Huang:2021nkl,Yin:2020afe,Buttazzo:2020eyl,Capdevilla:2020qel,Han:2020pif,Han:2020uak,Costantini:2020stv}. Recent studies indeed demonstrated the impressive physics potentials exploring the EW sector, including precision Higgs boson coupling measurements \cite{Han:2020pif}, the electroweak dark matter detection \cite{Han:2020uak}, and discovery of other BSM heavy particles \cite{Costantini:2020stv}.

In this article, we explore the discovery potential for the non-SM heavy Higgs bosons at a high-energy muon collider in the framework of 2HDMs \cite{Branco:2011iw}. We adopt the commonly studied four categories according to the assignments of a discrete $\mathbb{Z}_2$ symmetry, which dictates the pattern of the Yukawa couplings. We identify the relevant parameters of the Higgs masses and couplings and predict the decay branching fractions. We take a conservative approach in the alignment limit for the mixing parameter so that there are no large corrections to the SM Higgs physics.

We consider the benchmark energies for the  muon colliders  \cite{Delahaye:2019omf} in the range of $\sqrt s=3-30$ TeV, with the integrated luminosity scaled as
\beq
\label{eq:luminosity}
\mathcal{L}= ({\sqrt s\over 10\ {\rm TeV}})^2 \times 10^4\ {\rm fb}^{-1}.
\eeq
We study both the heavy Higgs boson pair production as well as single production associated with two heavy fermions. Both $\mm$ annihilation channels and Vector Boson Fusion (VBF) processes are considered, which are characteristically different. We also  analyze the radiative return $s$-channel production of a heavy Higgs boson in $\mm$ annihilation, given the possible enhancement of the muon Yukawa couplings in certain models. We further consider the Higgs bosons to decay to heavy fermions and study their signatures and the SM backgrounds. We design appropriate cuts to select the signals according to the different final states and kinematics, while effectively suppress the backgrounds.  Combining together the production channels and the decay patterns, we also show how the four different types of 2HDMs can be distinguished.

The rest of the paper is organized as follows.
In Sec.~\ref{sec:model}, we briefly introduce the 2HDMs, identify the relevant parameters of the Higgs masses and couplings and predict the decay branching fractions. In Sec.~\ref{sec:pair}, we present the results for the heavy Higgs boson pair production and discuss the signal observability above the SM backgrounds. We also discuss how different types of 2HDMs can be distinguished by studying the production and decays of heavy Higgs bosons. In Sec.~\ref{sec:fermion}, we present the results for single heavy Higgs boson production in association with a pair of fermions.  In Sec.~\ref{sec:radiative}, we present the radiative return production of a non-SM Higgs boson and compare that to the other production mechanisms. In Sec.~\ref{sec:conclusion}, we summarize and draw our conclusion.

%%%%%%%%%%%%%%%%%%%%%%%%%%%%%%%%
\section{Two Higgs doublet models}
\label{sec:model}

%%%%%%%%%%%%%%%%%%%%%%%%%%%%%%%%
\subsection{Higgs boson couplings}
\label{sec:intro}

The Higgs sector of the 2HDMs \cite{Branco:2011iw} consists of two SU(2)$_L$ scalar doublets  $\Phi_i\,(i=1,2)$ with hyper-charge $Y=1/2$
\begin{align}
        \Phi_i = \left(\begin{array}{c}
                \phi_i^+\\
                (v_i+\phi_i^0+iG_i^0)/\sqrt{2}
        \end{array}\right),
\end{align}
where $v_i\,(i=1,2)$ are the vacuum expectation values (vev) of the doublets after the electroweak symmetry breaking (EWSB), satisfying $\sqrt{v_1^2+v_2^2}=v=246 \GeV$.

After the electroweak symmetry breaking,
the scalar sector of the 2HDMs~\cite{Branco:2011iw} consists of 5 physical scalars:  $h,H,A,H^\pm$ with masses $m_h, m_H, m_A$ and $m_{H^\pm}$. In this work, $h$ will be identified as the observed SM-like Higgs with $m_h=125\,\rm GeV$, while all other scalars are heavy.   The tree-level couplings of Higgs bosons are determined by two parameters: the mixing angle between the neutral CP-even Higgs bosons $\alpha$ and $\tan\beta=v_2/v_1$.

There are three types of couplings between Higgs bosons and gauge bosons: the Gauge boson-Gauge boson-Scalar couplings, the Gauge boson-Scalar-Scalar couplings and the Gauge boson-Gauge boson-Scalar-Scalar couplings, all originated from the kinetic term of the Higgs fields.  The corresponding terms in the Lagrangian are expressed as
\begin{equation}
\begin{aligned}
\mathcal{L}\supset &\frac{1}{s_{\phi V_1V_2}}g_{\phi V_1V_2}g^{\mu\nu}\phi V_{1\mu} V_{2\nu}+\frac{1}{s_{\phi_1\phi_2 V}}g_{\phi_1\phi_2 V}(\partial^\mu\phi_1 \phi_2-\phi_1\partial^\mu\phi_2)V_\mu\\
&+\frac{1}{s_{\phi_1\phi_2 V_1V_2}}g_{\phi_1\phi_2 V_1V_2}g^{\mu\nu}\phi_1\phi_2 V_{1\mu}V_{2\nu}\  ,
\end{aligned}
\end{equation}
where $s_{\phi V_1V_2}$, $s_{\phi_1\phi_2 V}$ and $s_{\phi_1\phi_2 V_1V_2}$ are the symmetry factors. %such as $1!, 2!, \cdots$.
The coupling strengths $g_{\phi V_1V_2}$, $g_{\phi_1\phi_2 V}$ and $g_{\phi_1\phi_2 V_1V_2}$ are summarized in \autoref{tb:guagehiggs} \cite{Kanemura:2015mxa}, where we keep the dependence on the mixing parameter $\sin(\beta-\alpha) \equiv s_{\beta-\alpha}$ and  $\cos(\beta-\alpha) \equiv c_{\beta-\alpha}$.

In this table, the interactions are classified into suppressed ($\propto c_{\beta-\alpha}$) and un-suppressed ($\propto s_{\beta-\alpha}$) categories. This is motivated by the scaling properties of the coupling strengths in the limit of $\cos(\beta-\alpha)=0$, the so-called \textit{alignment limit}. Under this limit, the couplings of the SM-like Higgs boson $h$ with two gauge bosons restore the couplings in the SM, while the couplings of pair of SM gauge bosons to the non-SM Higgs bosons vanish.  Given the Higgs coupling measurements at the LHC~\cite{Gu:2017ckc,Su:2019ibd}, the 2HDM parameter spaces have already been constrained to be near the
alignment limit~\cite{Chen:2018shg,Chen:2019pkq,Chen:2019bay,Chen:2019rdk} .  In our analyses below, we will assume the alignment limit of $\cos(\beta-\alpha)=0$ when $h$ will have   the same tree-level couplings to gauge bosons and fermions as in the SM, and the other heavy Higgs bosons exhibit the universal gauge interactions. Even in the large $\tan\beta$ region of the Type-I 2HDM when the largest deviation of $\cos(\beta-\alpha)$ from 0 is possible (at tree-level): $\cos(\beta-\alpha)\lesssim 0.3$, the production cross sections will not be suppressed too much comparing to the results under the  exact alignment limit as we presented in this paper.  Furthermore, additional  production channels could appear for non-zero $\cos(\beta-\alpha)$, for example, $\mu^+\mu^- \rightarrow hA$, which could be used to extend the reach of BSM Higgs bosons beyond the pair production threshold.

Note that for $\phi_1\phi_2 V$ and $\phi_1\phi_2 V_1 V_2$ couplings, interactions involving two non-SM Higgs bosons are un-suppressed under the alignment limit, while interactions involving one SM-like Higgs $h$ and one non-SM Higgs are always suppressed.  This leads to the pair production of non-SM Higgs bosons  via the full gauge interaction strength at a muon collider, while the single non-SM Higgs production in association with a SM gauge boson is suppressed by the small mixing.

\begin{table}[h]
\begin{center}
\begin{tabular}{ r | c l | c  l | c l }
\hline
& \multicolumn{2}{|c|}{$\phi V_\mu V_\nu$ }& \multicolumn{2}{|c|}{$\phi_1\phi_2 V_\mu$ } & \multicolumn{2}{|c}{$\phi_1\phi_2 V_{1\mu} V_{2\nu}$ }\\
Factor & \multicolumn{2}{|c|}{$ig_{\mu\nu}$ }& \multicolumn{2}{|c|}{$(p_1-p_2)_\mu$ } & \multicolumn{2}{|c}{$ig_{\mu\nu}$ }\\
\hline
\multirow{14}{*}{leading }&$ hW^+W^-$& $\frac{g^2v}{2}s_{\beta-\alpha}$& $HH^\pm W^\mp$ &$\pm i\frac{g}{2}s_{\beta-\alpha}$& $hhW^+W^-$ & $\frac{g^2}{2}$ \\
& & & $AH^\pm W^\mp$ & $-\frac{g}{2}$ & $HHW^+W^-$ & $\frac{g^2}{2}$\\
 & & & & & $AAW^+W^-$ & $\frac{g^2}{2}$\\
 & & & & & $H^+H^-W^+W^-$ & $\frac{g^2}{2}$\\
 \cline{2-7}
 & $ hZZ$& $\frac{g^2v}{2c_W^2}s_{\beta-\alpha}$ & $H^+H^-Z$ & $i\frac{g}{2c_W}c_{2W}$ & $hhZZ$ & $\frac{g^2}{2c_W^2}$\\
 & & & $HAZ$ & $\frac{g}{2c_W}s_{\beta-\alpha}$  & $HHZZ$ & $\frac{g^2}{2c_W^2}$\\
 & & & & & $AAZZ$ & $\frac{g^2}{2c_W^2}$\\
 & & & & & $H^+H^-ZZ$ & $\frac{g^2}{2c_W^2}c_{2W}^2$\\
 & & & & & $H^\pm HW^\mp Z$ & $\frac{g^2}{2c_W}s_W^2 s_{\beta-\alpha}$\\
 & & & & & $H^\pm AW^\mp Z$ & $\pm i\frac{g^2}{2c_W}s_W^2$\\
  \cline{2-7}
 & & & $H^+H^-\gamma$ & $ie$ & $H^+H^-\gamma\gamma$ & $2e^2$\\
 & & & & & $H^\pm AW^\mp \gamma$ & ${\mp i}\frac{eg}{2}$\\
 & & & & & $H^+ H^- Z \gamma$ & $\frac{eg}{c_W}c_{2W}$\\
 & & & & & $H^\pm H W^\mp \gamma$ & $-\frac{eg}{2}s_{\beta-\alpha}$\\
\hline
\multirow{2}{*}{suppressed }&$ HW^+W^-$& $\frac{g^2v}{2}c_{\beta-\alpha}$& $hH^\pm W^\mp$ & $\mp i\frac{g}{2}c_{\beta-\alpha}$ & $H^\pm h W^\mp \gamma$ & $\frac{eg}{2}c_{\beta-\alpha}$\\
&$ HZZ$& $\frac{g^2v}{2c_W^2}c_{\beta-\alpha}$& $hAZ$ & $-\frac{g}{2c_W}c_{\beta-\alpha}$ &  $H^\pm h W^\mp Z$& $-\frac{g^2}{2c_W}s_W^2c_{\beta-\alpha}$\\
\hline
\end{tabular}
\end{center}
\caption{Coupling strengths of $g_{\phi V_1V_2}$, $g_{\phi_1\phi_2 V}$ and $g_{\phi_1\phi_2 V_1V_2}$ in the 2HDMs for $V=W,Z, \gamma$, $\phi=h,H,H^\pm,A$, where $s_W^{}$ ($c_W^{}$) is the sine (cosine) of the weak mixing angle $\theta_W$, and $c_{2W}\equiv \cos(2\theta_W)$,  $s_{\beta-\alpha} \equiv \sin(\beta-\alpha)$ and  $\cos(\beta-\alpha) \equiv c_{\beta-\alpha}$.}
\label{tb:guagehiggs}
\end{table}

The Lagrangian of Yukawa couplings is
\begin{equation}
  -\mathcal{L}_{\rm Yuk}=Y_{d}{\overline Q}_L\Phi_d d_R^{}+Y_{e}{\overline L}_L\Phi_e e_R^{}+ Y_{u}{\overline Q}_L\tilde\Phi_u u_R^{}+\text{h.c.}\, ,
  \label{eq:yukawa}
\end{equation}
where $\tilde{\Phi}=i\sigma_2\Phi^*$ and $\Phi_{u,d,e}$ are either $\Phi_{1}$ or $\Phi_{2}$, depending on the $\mathbb{Z}_2$ charge assignments.
Expanding~\autoref{eq:yukawa} in terms of the physical Higgs fields, the interactions of Higgs bosons with fermions can be expressed as
\begin{equation}
\begin{aligned}
\mathcal{L}_{\rm Yuk}=&-\sum_{f=u,d,\ell}\frac{m_f}{v}\biggl(\xi_{Hff}\bar f fH-i\xi_{Aff}\bar f\gamma_5 fA\biggr)\\
&-\biggl\{\frac{\sqrt{2}V_{ij}}{v}\bar u_i(m_{u_i}\xi_{Auu} P_L+m_{d_j}\xi_{Add} P_R)d_jH^+ +\frac{\sqrt{2}m_\ell\xi_{A\ell}}{v}\bar \nu_L\ell_RH^++h.c.\biggr\} ,
\end{aligned}
\end{equation}
where $u= (u,c,t)$, $d=(d,s,b)$, $V_{ij}$ is the CKM matrix, and $P_{L/R}\equiv (1\mp\gamma^5)/2$ are the projection operators for the left-/right-handed fermions. In this expression, factors $\xi$ are the  coupling strengths normalized to the corresponding SM value, which are shown below at \autoref{eq:normcoupling} under the alignment limit of  $\cosba=0$.

\begin{table}[tb]
    \begin{center}
        \begin{tabular}{ccccccccc}
            \hline\hline
            Types   & $\Phi_1$ & $\Phi_2$ & $u_R$ & $d_R$ & $\ell_R$ & $Q_L$, $L_L$&$\Phi_1$ & $\Phi_2$ \\
            \hline
            Type-I & $+$ & $-$ & $-$ & $-$ & $-$ & $+$&&$u$, $d$, $\ell$ \\
            Type-II & $+$ & $-$ & $-$ & $+$ & $+$ & $+$ &$d$, $\ell$& $u$\\
            Type-L & $+$ & $-$ & $-$ & $-$ & $+$ & $+$ & $\ell$&$u$, $d$,\\
            Type-F & $+$ & $-$ & $-$ & $+$ & $-$ & $+$ &$d$&$u$, $\ell$\\
            \hline\hline
        \end{tabular}
        \caption{Four types of  assignments for the $\mathbb{Z}_2$ charges and the Yukawa interactions  for the scalar doublets $\Phi_{1,2}$ and the SM fermions.}
        \label{tab:z2_assign}
    \end{center}
\end{table}

There are four different types of Yukawa coupling structure depending on how the two Higgs doublets are coupled to the leptons and quarks.  The $\mathbb{Z}_2$ charge assignments and the corresponding Yukawa interactions are listed in~\autoref{tab:z2_assign}.   The  tree-level expressions of various $\xi$s  can be found in Ref.~\cite{Branco:2011iw}.  Under the alignment limit of $\cos(\beta-\alpha)=0$,  the Higgs couplings of $H$ and $A$ to the SM fermions normalized to the SM values are
\begin{align}
 &\textbf{Type-I:}  &\xi_{Huu}=\xi_{Auu}&=\cot\beta,  &\xi_{Hdd}=-\xi_{Add}&=\cot\beta, &\xi_{H\ell\ell}=-\xi_{A\ell\ell}&=\cot\beta; \nonumber\\
 &\textbf{Type-II:}  &\xi_{Huu}=\xi_{Auu}&=\cot\beta,  &-\xi_{Hdd}=\xi_{Add}&=\tan\beta, &-\xi_{H\ell\ell}=\xi_{A\ell\ell}&=\tan\beta; \nonumber\\
 &\textbf{Type-L:}  &\xi_{Huu}=\xi_{Auu}&=\cot\beta,  &\xi_{Hdd}=-\xi_{Add}&=\cot\beta, &-\xi_{H\ell\ell}=\xi_{A\ell\ell}&=\tan\beta; \nonumber\\
 &\textbf{Type-F:}  &\xi_{Huu}=\xi_{Auu}&=\cot\beta,  &-\xi_{Hdd}=\xi_{Add}&=\tan\beta, &\xi_{H\ell\ell}=-\xi_{A\ell\ell}&=\cot\beta.
 \label{eq:normcoupling}
\end{align}
While the top-quark Yukawa coupling is always enhanced at small $\tan\beta$ and suppressed at large $\tan\beta$, the Yukawa  couplings to the bottom quark and tau lepton can be either suppressed or enhanced in difference regions of $\tan\beta$, depending on the types of 2HDMs.

Theoretical considerations, such as the requirements of vacuum stability,  perturbativity and unitarity impose additional constraints on the 2HDM model parameters, which can be translated to the masses and their splittings of the non-SM  Higgs bosons. Detailed analyses can be found in Refs.~\cite{Ginzburg:2005dt,Kanemura:2015ska,Goodsell:2018fex,Barroso:2013awa,Xu:2017vpq}.  The constraints are the weakest for $\lambda v^2 = m_H^2 - m_{12}^2/(s_\beta c_\beta)=0$ under the degenerate mass assumptions, when all values of $\tan\beta$ are allowed.  The allowed range of $\tan\beta$ shrinks when the value of $\lambda v^2$ deviates from zero. For example, for $\lambda v^2 = (300~{\rm GeV})^2$, the allowed range of $\tan\beta$ is between  $0.4$ and  $2.5$ under the alignment and degenerate mass limits~\cite{Gu:2017ckc}. Furthermore, ranges of the scalar mass splittings also depend on the values of $\lambda v^2$.  For our numerical  analyses of non-SM Higgs production below, we set $\lambda v^2=0$.  Non-zero values of $\lambda v^2$ will not change the cross sections significantly.

There have been extensive searches for non-SM Higgs bosons at the LHC~\cite{Kling:2020hmi,Su:2019ibd}.  For the neutral Higgs bosons, $A/H\rightarrow \tau^+\tau^-$ sets the strongest direct search limits.  For Type-II 2HDM with enhanced $bb$ and $\tau\tau$ Yukawa
couplings at large $\tan\beta$, the lower limit on $m_{A/H}$ is 1 TeV (1.5 TeV) for $\tan\beta=10$ (50)~\cite{Kling:2020hmi,Su:2019ibd}.  At the small $\tanb$ region, both Type-I and Type-II get strong constraints from $A/H\to \gamma\gamma$, $A/H \to t\bar t$ and four tops non-resonance search channels.  For $\cosba\sim 0$, $m_{H/A}<4 m_t$ with $\tanb<1$ is currently excluded, and the exclusion reaches 1 TeV for $\tanb<0.2$.

Electroweak precision measurements also provide indirect constraints on the mass splitting between $H/A$ and $H^\pm$.  Given the current 95\% C.L. range of the oblique parameters~\cite{Chen:2018shg,Chen:2019pkq}, the charged Higgs boson mass is constrained to be close to one of the neutral Higgs  masses: $m_{H^\pm} \sim m_{H}$ or  $m_{H^\pm} \sim m_{A}$. Precision measurements at the $Z$-pole and the Higgs factory could further limit the range of the mass splittings, as studied in Refs.~\cite{Chen:2018shg,Chen:2019pkq}. In addition, since the mass differences between the BSM Higgses are controlled by the quartic couplings in the Higgs potential, theoretical considerations (vacuum stability, unitarity and perturbativity) result in $\Delta m \lesssim 200\  (100)$ GeV   for $m_\Phi = 1\ (2) $ TeV~\cite{Kling:2018xud}.   In our analyses below, we take the degenerate mass assumption of $m_\Phi\equiv m_H=m_{A}=m_{H^\pm}$.  Numerically, the pair production cross of $H^+H^-$  or Higgs produced in associated with a pair of heavy fermions via annihilation will not change since it only depends on the corresponding BSM Higgs mass.  Other production process could have different production cross sections,  due to the change of the masses of either the intermediate or the final state Higgses.  The numerical difference, however, is not expected to be large given the viable ranges of the mass differences mentioned above.   It is also worth mentioning that once the mass difference is larger than $m_{W,Z}$, additional exotic decay modes open, for example, $A\rightarrow HZ$, $H^\pm \rightarrow HW^\pm$, which could be used to enhance the reach~\cite{Kling:2018xud}.

Flavor constraints, such as those from  $B_d^0-\bar{B}_d^0$ mixing, $b\rightarrow s \gamma$, $B$ and $D$ meson/baryon decays also bound the 2HDM parameter space, in particular, on $m_{H^\pm}$ and regions with Yukawa coupling enhancement. Most notably, the measured value for BR($b \to s \gamma$) imposes a lower limit on the charged Higgs mass to be larger than  $\sim 600~\gev$ in the Type-II 2HDM~\cite{Haller:2018nnx,Amhis:2016xyh}.

%%%%%%%%%%%%%%%%%%%%%%%%%%%%%%%%%%%
\subsection{Higgs boson decays}
\label{sec:decay}

Under the alignment limit with degenerate heavy Higgs boson masses, only the decays of the heavy Higgs bosons to a pair of SM fermions are relevant. Since the couplings to the fermions are proportional to their masses, the leading decay channels are to the heavy fermions
\beq
H/A \to t\bar t,\ b\bar b,\ {\rm and}\ \tau^+\tau^-,\quad
H^\pm \to t  b,\ {\rm and}\ \tau \nu_\tau.
\label{eq:decay}
\eeq

\begin{figure}[tb]
\centering
\includegraphics[width=\textwidth]{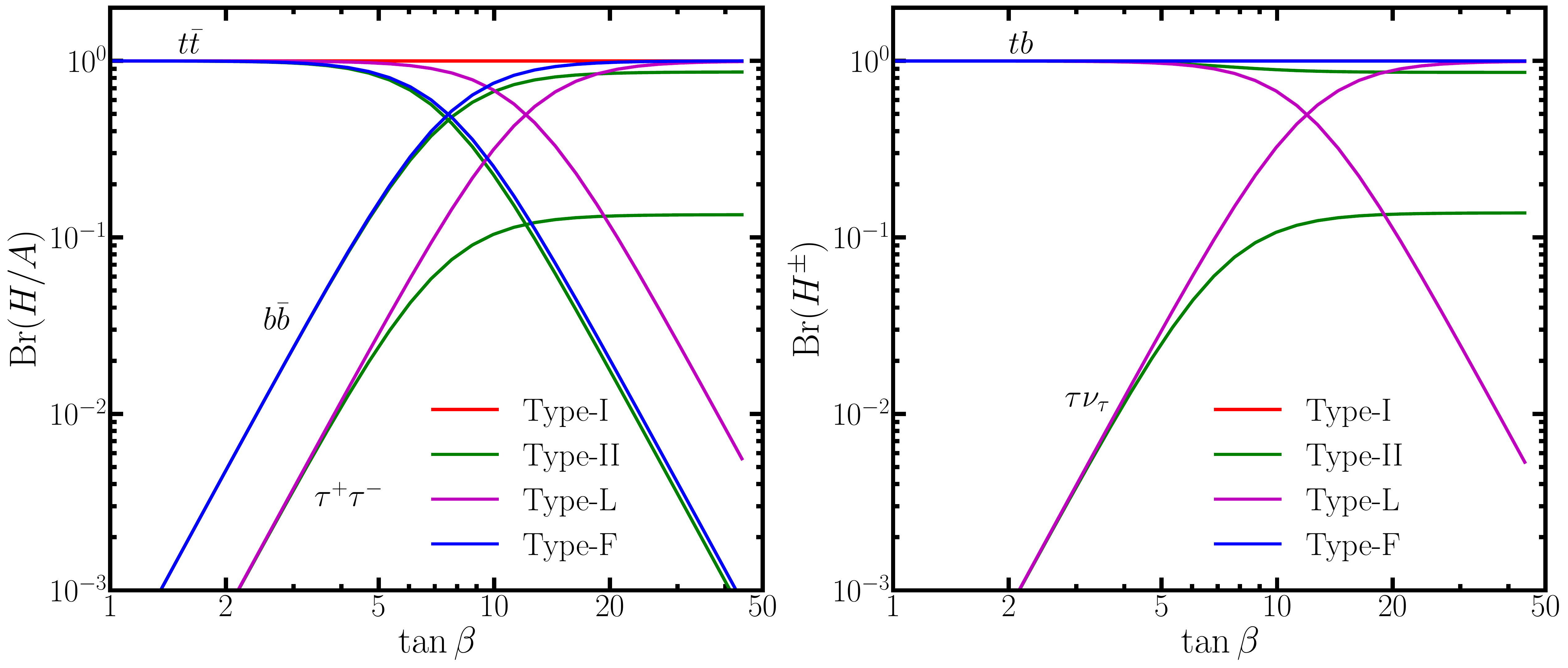}
\caption{Leading decay branching fractions of $H(A)$ (left panel) and $H^\pm$ (right panel) as a function of $\tan\beta$ in four Types of 2HDMs for $m_\Phi=2$ TeV and $\cos(\beta-\alpha)=0$.
In the right panel, red curve (Type-I) overlaps with the blue curve (Type-F). }
\label{fig:BR}
\end{figure}

For illustration, we take  $m_\Phi\equiv m_H=m_{A}=m_{H^\pm}=2$ TeV, under the  alignment limit $\cos(\beta-\alpha)=0$.
We calculate the decay branching fractions in all the four types of 2HDMs, using {\tt 2HDMC}~\cite{Eriksson:2009ws}, and the results are presented in~\autoref{fig:BR}.
The decay channels $t\bar t$, $b\bar b$, $\tau^+\tau^-$ for $H/A$ (left panel), and $tb$,  $\tau\nu_\tau$ for $H^\pm$ (right panel) are shown separately. In the limit of $m_f\ll m_\Phi$, $H$ and $A$ have identical branching fractions.

The branching fractions of the three leading decay channels exhibit an apparent hierarchical behavior due to the non-universal Yukawa couplings: the $t\bar t$ decay for $H/A$ ($tb$ for $H^\pm$) always dominates except when there are strong enhancements of other decay channels at large $\tan\beta$. For example,  $H/A \rightarrow b\bar b$ dominates in the large $\tan\beta$ region for Type-II and F, and $H/A \rightarrow \tau^+\tau^-$ dominates in the large $\tan\beta$ region for Type-L. For the charged Higgs $H^\pm$ decay, the suppression of $tb$ decay caused by the enhancement of other decay channels is less obvious given that $H^\pm \rightarrow tb$  also enhances at large $\tan\beta$  in both Type-II and Type-F. $H^\pm \rightarrow \tau\nu_\tau$, however, could be dominant in the large $\tan\beta$ region for Type-L.

One noteworthy point is, while the decay branching fraction of the dominant decay mode for the different types of 2HDMs degenerates at small $\tan\beta$, it is quite distinct at large $\tan\beta$, which would allow the discrimination between different types of 2HDMs by examining the decays of heavy Higgs bosons.

%%%%%%%%%%%%%%%%%%%%%%%%%%%%%%%%%%%%%%%%%%%%%%%%
\section{Higgs pair production in $\mu^+\mu^-$ annihilation and vector boson fusion}
\label{sec:pair}

A high energy muon collider would have the capability to open a new energy threshold at the energy frontier. While the $\mm$ annihilation will be most efficient in exploiting the available c.m.~energy for heavy particle production, it has been argued that the VBF processes will become increasingly more important at higher energies and offer a variety of production channels due to the initial state spectrum.

%%%%%%%%%%%%%%%%%%%%%%%%%%%%%%
\subsection{Production cross sections}

\begin{figure}[!tbp]
    \centering
    \includegraphics[width=0.3\textwidth]{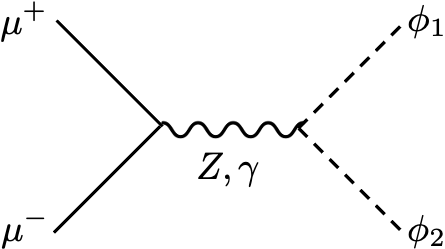}
    \caption{Representative Feynman diagram for the EW scalar pair production in $\mm$ annihilation $\mm\to \phi_1 \phi_2 $.  }
    \label{fig:Feyn_ffss}
\end{figure}

\begin{figure}[tb]
\centering
\includegraphics[width=\textwidth]{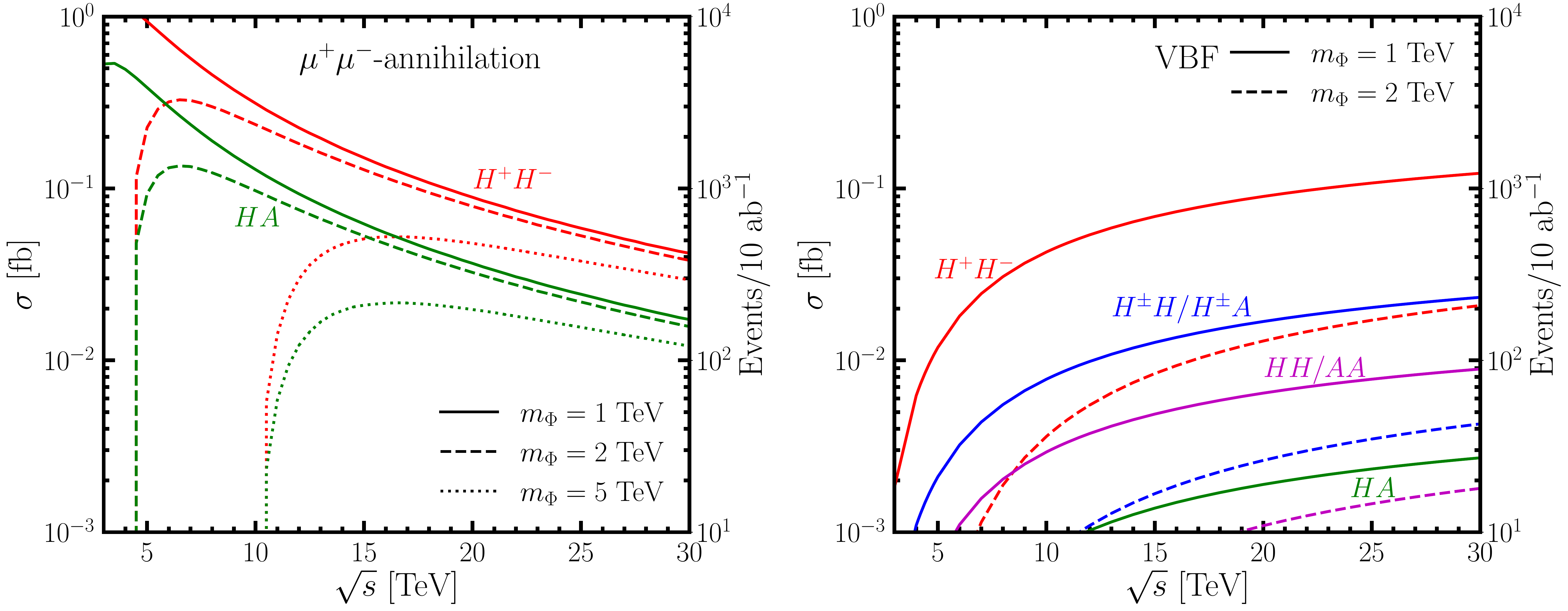}
\caption{Cross sections versus the  c.m.~energy $\sqrt{s}$. For the left panel: $\mu^+\mu^-\to H^+H^- $ (red), and $HA$ (green) through $\mu^+\mu^-$ annihilation; and for the right panel: in addition, $H^\pm H/H^\pm A$ (blue), $HH/AA$ (purple), through VBF in the alignment limit $\cos(\beta-\alpha)=0$. Solid, dashed and dotted lines for degenerate Higgs masses $m_\Phi=1$ TeV, 2 TeV and 5 TeV, respectively. The vertical axis on the right shows the corresponding event yields for a 10 ab$^{-1}$ integrated luminosity.  }
\label{fig:Pair}
\end{figure}

Once crossing the pair production threshold, the heavy Higgs bosons can be produced in pair via the $\mm$ annihilation
\begin{equation}
\mm\to \gamma^*,Z^*\to H^+ H^-,\quad
\mm \to Z^* \to HA.
\label{eq:anni}
\end{equation}
The Feynman diagrams of the leading contributions are shown in~\autoref{fig:Feyn_ffss}.   In the alignment limit  of $\cos(\beta-\alpha)=0$,
%and $\lambvs=0$,
the production is fully governed by the EW gauge interactions, which are universal for all types of the 2HDMs. The left panel of~\autoref{fig:Pair} shows the total cross sections of~\autoref{eq:anni} versus the collider c.m.~energy $\sqrt s$ for
degenerate heavy Higgs masses $m_\Phi(=m_H=m_A=m_{H^\pm})=$1 TeV (solid curves), 2 TeV (dashed curves) and 5 TeV (dotted curves).     Red and green curves are   for $H^+H^-$ and $HA$ productions.
 We see the threshold behavior for a scalar pair production in a P-wave as $\sigma \sim \beta^3$, with $\beta = \sqrt{1-4m_H^2/s}$. Well above the threshold, the cross sections asymptotically approach $\sigma \sim  \alpha^2/s$, which is insensitive to the heavy Higgs mass. The excess of the $H^+H^-$ production cross section over that of $HA$ is attributed to the $\gamma^*$-mediated process. The vertical axis on the right shows the corresponding events for a 10 ab$^{-1}$ integrated luminosity, yielding a sizeable event sample. The cross sections are calculated using {\tt MadGraph5 V2.6.7} \cite{Alwall:2014hca} with Initial State Radiation (ISR) accounted \cite{Li:2018qnh}.

\begin{figure}
    \centering
    \includegraphics[width=\textwidth]{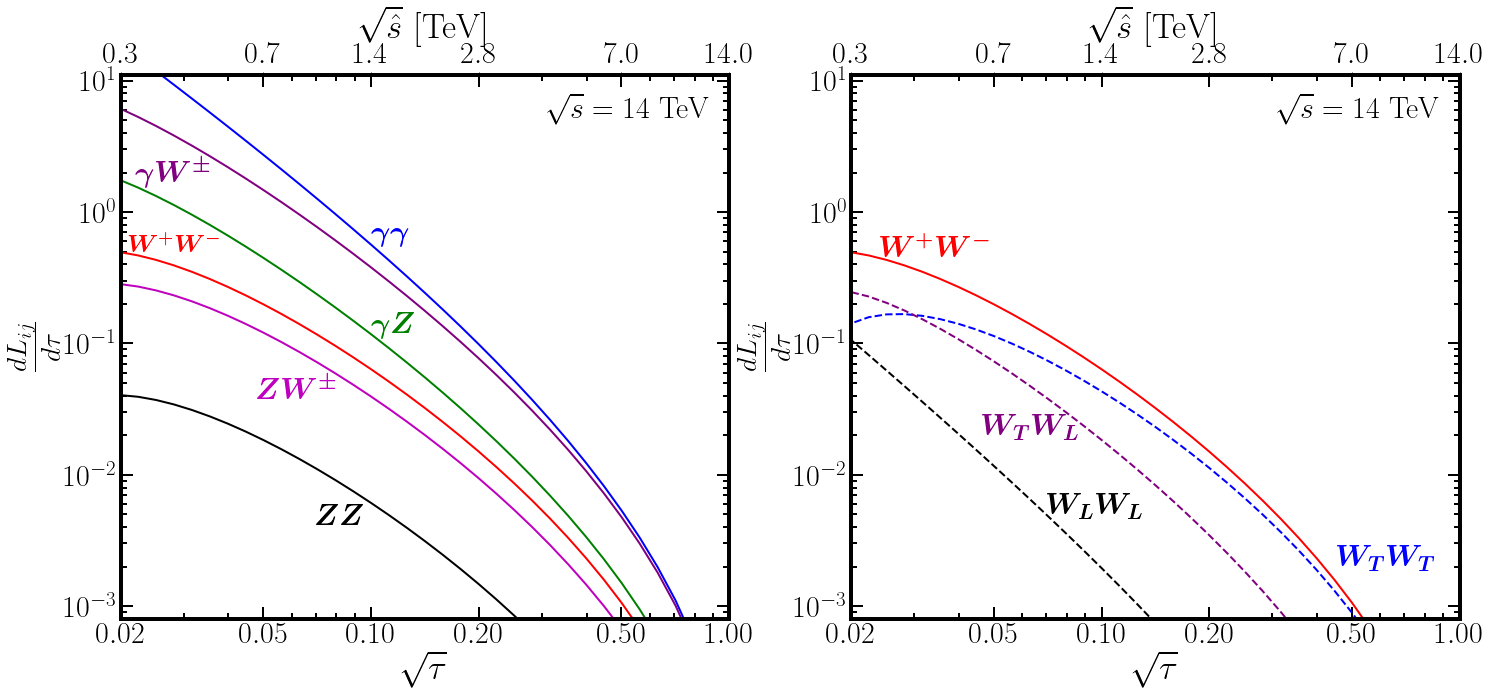}
    \caption{Vector boson parton luminosities  $d\mathcal{L}_{ij}/{d\tau}$ versus  $\tau =\hat{s}/s$ at  $Q=\sqrt{\hat{s}}/2$ and $\sqrt s= 14$ TeV.}
    \label{fig:parton_luminosity}
\end{figure}

At high c.m.~energies, the VBF processes become increasingly important. For a fusion process of the initial state vector boson partons $V_i$ and $V_j$, we write the production cross section of an exclusive final state $F$ and the unspecified remnants $X$
in terms of the parton luminosity $d\mathcal{L}_{ij}/{d\tau}$ and the corresponding partonic sub-process cross section $ \hat{\sigma}$
\begin{eqnarray}
\label{eq:crosssection}
&  \sigma(\ell^+\ell^- \rightarrow F + X)
= \int_{\tau_{0}}^{1} d\tau  \sum_{ij}\frac{d\mathcal{L}_{ij}}{d\tau}\  \hat{\sigma}(V_i V_j\rightarrow F),
\nonumber \\
& { d\mathcal{L}_{ij} \over {d\tau} } = \frac{1}{1+\delta_{ij}}  \int^{1}_{\tau} \frac{d\xi}{\xi}
 \left[ f_{i}(\xi, Q^{2})f_{j}\left(\frac{\tau}{\xi},Q^{2} \right) + (i \leftrightarrow j) \right],
\label{eq:fact}
\end{eqnarray}
where
$\tau = \hat s/s$ with  $\sqrt{\hat{s}}$ being  the  parton-level  c.m.~energy.   The production threshold is $\tau_{0} = m_{F}^{2}/s$. In this expression, $f_i(\xi,Q^2)$ stands for the electroweak parton distribution function (EW PDF) of particle $V_i$ radiated off the initial muon beam carrying an energy fraction $x$ at a factorization scale $Q$. In our study we adopt the leading-order
EW PDFs \cite{DAWSON198542,Kane:1984bb}.
Recently the EW PDFs have been calculated with the DGLAP evolution at the double-log accuracy \cite{Han:2020uid}. The numerical difference from  the leading-order result is typical less than $10\%$.
In~\autoref{fig:parton_luminosity}, we present the vector-boson luminosity $d\mathcal{L}_{ij}/{d\tau}$  for $\mu^+\mu^-$ collisions at 14 TeV. The QED $\gamma\gamma$ luminosity in the left panel is the largest at the low $\hat s$ from the enhancement of $\log(Q^2/m_\mu^2)$ versus $\log(Q^2/m_V^2)$ for massive vector bosons, while the difference becomes much smaller at higher energies. Such logarithmic enhancements are absent for the longitudinal massive gauge bosons, which leads to the   suppressed partonic luminosity, as seen for $W_LW_L$ luminosity by the dashed black curve in the right panel. The smallness of the $ZZ$ luminosity is related to the accidentally small vector-like coupling of the neutral current, proportional to $(1/2-2\sin^2\theta_W)$ for the un-polarized muon beam.

\begin{figure}
    \centering
    \includegraphics[width=1.0\textwidth]{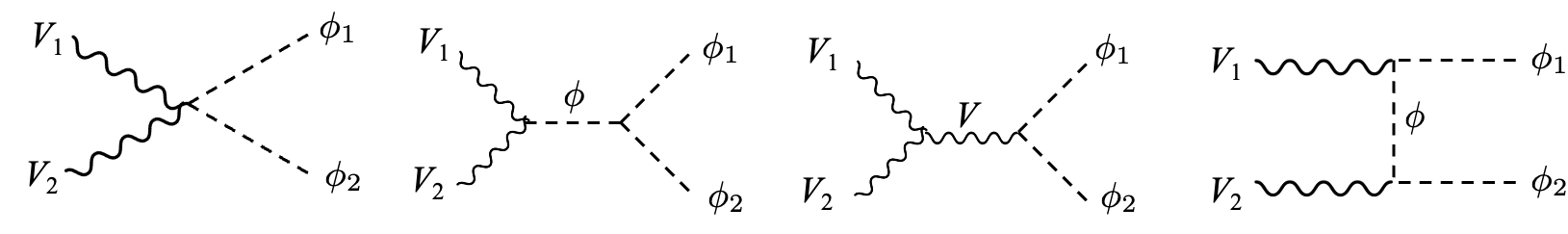}
    \caption{Representative Feynman diagrams for the VBF process $V_1 V_2\to \phi_1 \phi_2$.   }
    \label{fig:Feyn_vvss}
\end{figure}

Heavy Higgs boson pair production via VBF is via
\begin{equation}
    \mu^+\mu^-\to V_1 V_2\ \mu^+(\bar\nu)\mu^-(\nu),\ \ V_1V_2\to H^+H^-,\ HA,\
    H^\pm H/H^\pm A,\  HH/AA,
\end{equation}
where $V_1V_2=\gamma\gamma, W^+W^-, ZZ, Z\gamma, W^\pm Z, W^\pm \gamma$. The Feynman diagrams for the dominant contributions are shown in ~\autoref{fig:Feyn_vvss}. The cross sections over c.m.~energy $\sqrt{s}$ are   shown in the right panel of~\autoref{fig:Pair} for $H^+H^-$ (red), $HA$(green), $H^\pm H/H^\pm A$ (blue), and $HH/AA$ (purple). We see the expected logarithmic enhancement over the energy $\log^2(s/m_\mu^2)$ (or $\log^2(s/m_V^2)$) for initial state photons (weak bosons). Unlike the production by the $\mu^+\mu^-$ annihilation,  the cross section for the VBF processes are very sensitive to the heavy Higgs masses. The decrease of the cross sections with large $m_\Phi$ is primarily from the suppression of EW PDF threshold $\sim 1/M_F^2$.
 Again, the dominance of $H^+H^-$ production cross section over the neutral ones is due to the exclusive contribution from the $\gamma\gamma$ fusion.
 $HA$ production cross section is much smaller comparing to other processes since
 the first diagram of 4-point interaction in \autoref{fig:Feyn_vvss} without a heavy propagator suppression gives the leading contribution, which is absent for the $HA$ production.

In general, the $\mm$ annihilation process yields more Higgs pairs than the VBF process, except for the $H^+H^-$ production, when VBF takes over from $\sqrt{s}>20$ TeV for $m_{H^\pm}=1$ TeV. The typical cross section at $\sqrt{s}=6$ TeV, 14 TeV, 30 TeV and $m_\Phi=1$ TeV, 2 TeV and 5 TeV are summarized in \autoref{tab:higgpair}. With the benchmark luminosity given in \autoref{eq:luminosity}, a high energy muon collider can generate $O(10^3)$ $HA$ events once crossed the threshold and $O(10^4)$ $H^+H^-$ events at $m_{H^\pm}=$1 TeV.

\begin{table}[tb]
    \centering
    \begin{tabular}{|c c|c c|c c|c|}
    \hline
    \hline
    \multicolumn{2}{|c|}{\multirow{2}{*}{$\sigma$ (fb)}} & \multicolumn{2}{|c|}{$H^+H^-$}  & \multicolumn{2}{|c|}{$HA$}  & $HH^\pm$  \\
     & & $\mu^+\mu^-$ & VBF & $\mu^+\mu^-$ & VBF & VBF \\
     \hline
     \hline
    \multicolumn{7}{|c|}{$\sqrt{s}=6$ TeV}\\
    \hline
    \multirow{3}{*}{$m_\Phi$ } & 1 TeV & 0.73 & $1.8\times 10^{-2}$ & 0.30 & $2.4\times 10^{-4}$ & $6.4\times 10^{-3}$ \\
     & 2 TeV & 0.32 & $5.5\times 10^{-4}$ & 0.13 & $1.7\times 10^{-6}$ & $2.2\times 10^{-4}$ \\
     & 5 TeV & 0 & 0 & 0 & 0 & 0 \\
    \hline
    \hline
    \multicolumn{7}{|c|}{$\sqrt{s}=14$ TeV}\\
    \hline
    \multirow{3}{*}{$m_\Phi$ } & 1 TeV & 0.17 & $6.4\times 10^{-2}$ & $7.0\times 10^{-2}$ & $1.3\times 10^{-3}$ & $2.4\times 10^{-2}$\\
     & 2 TeV & 0.14 & $7.4\times 10^{-3}$ & $5.9\times 10^{-2}$ & $1.4\times 10^{-4}$ & $3.0\times 10^{-3}$  \\
     & 5 TeV & $4.7\times 10^{-2}$ & $7.1\times 10^{-5}$ & $2.0\times 10^{-2}$ & $2.0\times 10^{-7}$ & $3.2\times 10^{-5}$  \\
    \hline
    \hline
    \multicolumn{7}{|c|}{$\sqrt{s}=30$ TeV}\\
    \hline
    \multirow{3}{*}{$m_\Phi$ } & 1 TeV & $4.2\times 10^{-2}$ & $0.12$ & $1.7\times 10^{-2}$ & $2.7\times 10^{-3}$ & $4.7\times 10^{-2}$ \\
     & 2 TeV & $3.8\times 10^{-2}$ & $2.1\times 10^{-2}$ & $1.6\times 10^{-2}$ & $5.2\times 10^{-4}$ & $8.5\times 10^{-3}$\\
     & 5 TeV & $2.9\times 10^{-2}$ & $1.1\times 10^{-3}$ & $1.2\times 10^{-2}$ & $2.3\times 10^{-5}$ & $4.9\times 10^{-4}$ \\
    \hline
    \end{tabular}
    \caption{Summary of Higgs pair production cross sections via $\mu^+\mu^-$ annihilation and VBF for $m_\Phi=$1, 2 and 5 TeV and $\sqrt s=6$, 14 and 30 TeV.}
    \label{tab:higgpair}
\end{table}

\begin{figure}[tb]
\centering
\includegraphics[width=\textwidth]{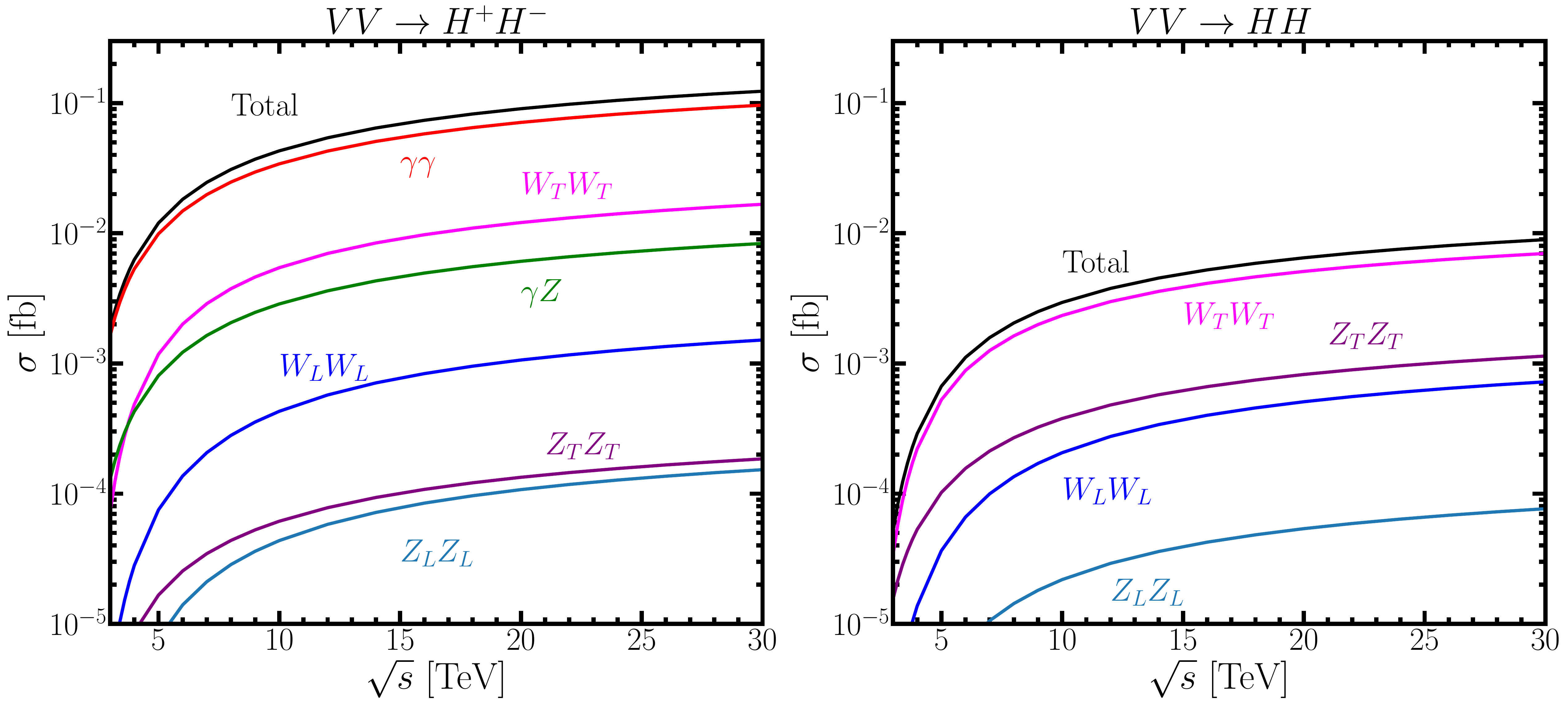}
 \caption{Cross sections versus the c.m.~energy $\sqrt s$ for individual contributions from different VBF  sub-processes to the production of $H^+H^-$ (left panel) and $HH$ (right panel), with a degenerate Higgs mass of $m_\Phi=1$~TeV.}
\label{fig:subprocess_pair}
\end{figure}

One of the advantages for adopting the EW PDF approach for
the calculations is the effective separation of the individual contributions from the fusion sub-processes. We illustrate this by presenting the contributing channels in~\autoref{fig:subprocess_pair} for $H^+H^-$ (left panel) and $HH$ (right panel) production.  We see that  the $\gamma\gamma$ (red) fusion sub-process contributes dominantly to the $H^+H^-$ production through electromagnetic interaction, which is explained by the abundant availability of $\gamma\gamma$ in~\autoref{fig:parton_luminosity}. Among the $WW$ fusions of different polarizations, $W_TW_T$ (purple) is much more copious than $W_LW_L$ (blue) because the scaling behavior of $\log(Q^2/m_W^2)$ for the transversely polarized gauge bosons is absent for the longitudinally polarized gauge bosons. The small contribution from the $ZZ$ fusion sub-process is related to its smallness in the partonic luminosity.   For the $HH$ production, all sub-process initiated with $\gamma$ fusion are absent, thus the production is mainly initiated by the $W_TW_T$ fusion and the total cross section is much smaller.  Compared to $H^+H^-$, the $ZZ$ fusion in $HH$ process is more prominent, which can be explained by the fact that the $ZH^+H^-$ coupling involved in $H^+H^-$ process is smaller than the $ZHA$ coupling involved in the $t$-channel contribution to $HH$ process by a factor of $c_{2W}$ (see~\autoref{tb:guagehiggs}).

\begin{figure}[tb]
\centering
\includegraphics[width=0.49\textwidth]{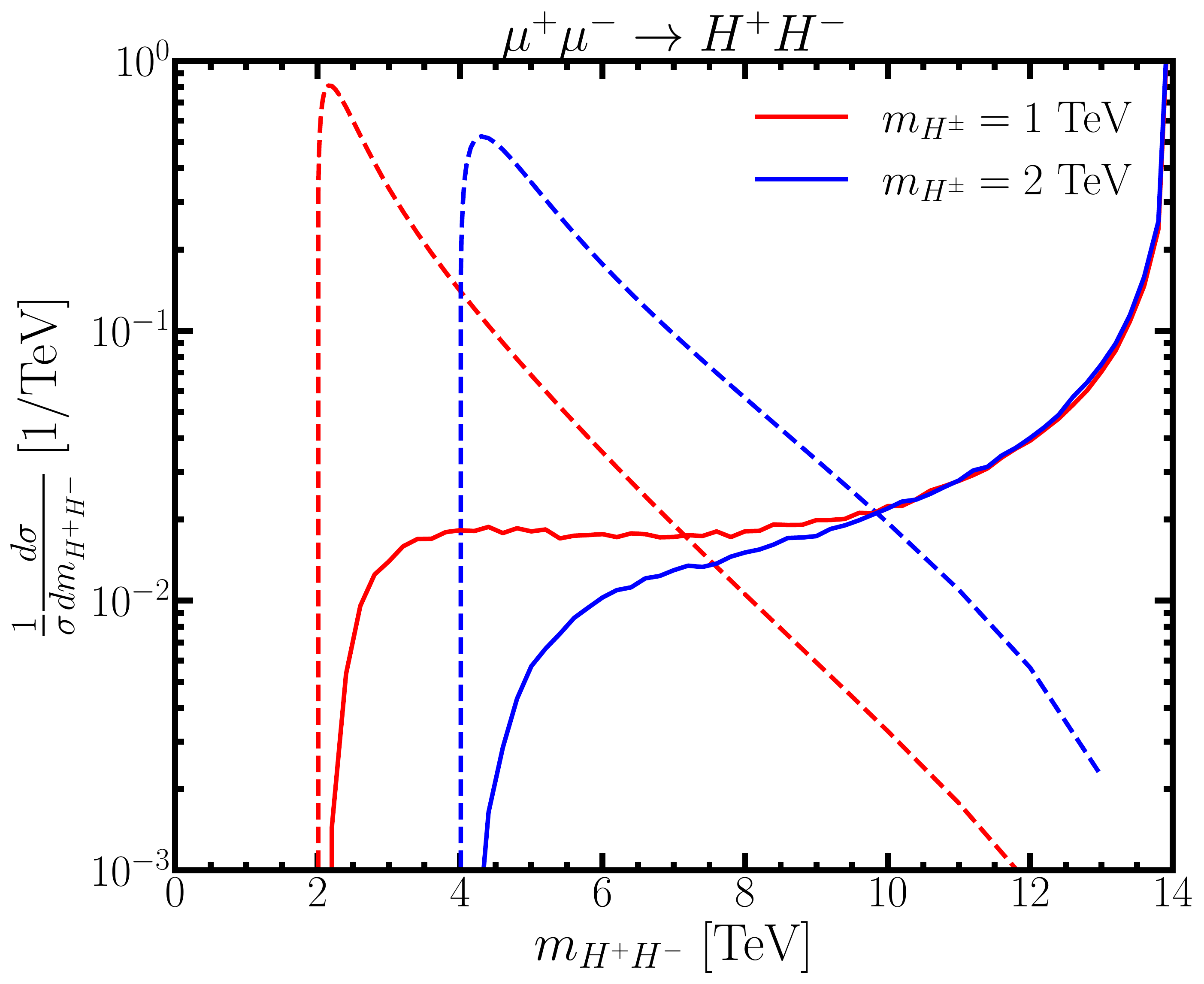}
\caption{Normalized invariant mass $m_{H^+H^-}$ distribution for $\mu^+\mu^-\to H^+H^- $  at $\sqrt{s}=14$ TeV for charged Higgs boson mass $m_{H^\pm}=1$ TeV (red) and 2 TeV (blue).  Solid (dashed) lines are for annihilation (VBF) contribution. }
\label{fig:Pair_distri}
\end{figure}

We note that, although a VBF channel has two accompanying leptons associated with the initial state gauge bosons, they are mostly unobservable due to their forward-backward collinear nature. As such, the $\mm$ annihilation and VBF both lead to the same observable Higgs pair final states. However, the invariant mass distributions of the Higgs pair system present a qualitatively different feature for these two processes, which may serve as an effective discriminator to separate these two processes.  The invariant mass of the Higgs pair is approximately equal to the collider c.m.~energy $m_{\phi_1\phi_2}\approx \sqrt{s}$ for the direct annihilation process, while peaked near the threshold $m_{\phi_1\phi_2}\approx m_{\phi_1}+m_{\phi_2}$ for the VBF process, as shown ~\autoref{fig:Pair_distri} for $H^+H^-$ production process. The long tail in the low invariant mass region for the annihilation process is due to the ISR effects taken into account in our calculations. For the rest of the analyses, we assume that the $\mm$ annihilation process and those from VBF are readily separable by kinematics.

%%%%%%%%%%%%%%%%%%%%%%%%%%%%%%
\subsection{Signals and backgrounds}
\label{sec:SB}

Given the pair-production of the heavy Higgs bosons and focusing on leading decay channels as shown in the last section, the signal will be four heavy third-generation fermions. Thus, the leading irreducible backgrounds are $t\bar t t\bar t$, $t\bar t b\bar b$ and $b\bar b b\bar b$ for third generation quark final states, dominated by the EW-QCD mixed processes of quark pair production, followed by a gluon radiation and splitting to another pair of quarks, typically $\mm \to t\bar t g^*, b \bar b g^*$ followed by $g^* \to b \bar b$.
For the sake of illustration, we consider the signal for a heavy Higgs mass $m_\Phi =2$ TeV and the corresponding backgrounds.
We note that the signal kinematics for the heavy Higgs pair production is quite different from the background processes, which would help for our signal reconstruction and background suppression.
First, the decay products of the heavy Higgs bosons possess a Jacobian peak in the transverse momentum $p_T \approx m_\Phi /2$ and they are more central in the angular distribution, while the fermions in the background tend to be softer and much more forward-backward.
We thus impose the basic acceptance cuts
\beq
p_T^t > 100~{\rm GeV},\ \ p_T^b> m_\Phi/5,\ \
10^\circ< \theta_{f} < 170^\circ,
\label{eq:basic_cuts_pair}
\eeq
where the choice for the angular cut is to simulate the detector coverage \cite{delphesTalk}.

Depending on the specific production channels and decays, the signals will be characterized by the mass reconstruction. We therefore further require
\bea
&& {\rm for}\ H^+H^-\ {\rm channel:}\quad
m_{tb}> 0.9 M_{H^\pm},\quad \theta_{tb} < 150^\circ ,\label{eq:HpHmcuts}\\
&& {\rm for}\ HA\ {\rm channel:}\quad
m_{tt},\ m_{bb}> 0.9 M_{H/A},\quad
\theta_{tt}, \theta_{bb} < 150^\circ.
\label{eq:HAcuts}
\eea
to reconstruct the resonance masses, where $\theta$ is the opening angle between the two fermions in the $\mm$ lab frame.
%. \Tao{(in which frame? in $\mm$ or partonic VBF?)}\Shuailong{In lab frame.} \Shufang{What's the different between theta(tb) and DeltaR(tb)? are these two equivalent?  In out cuts, some of them we use theta, some use DeltaR.} \Shuailong{Almost no difference by choosing either of them.}
The angular cut above is to suppress the background processes with the fermion pair back-to-back production, while the Higgs decay products are more collinear due to the potentially boosted Higgs bosons.
With those selection cuts above, the four heavy fermion backgrounds are highly suppressed as shown in \autoref{tab:pair_bg_xs}, resulting in the background rates about  $10^{-3}$ fb or below,
while the signal efficiencies shown in \autoref{tab:pair_sg_rate} are kept at $60\% - 80\%$, depending on the collider energy. Here we only calculated the signal efficiency for the $\mu^+\mu^-$-annihilation process since the production through VBF is much smaller at $\sqrt{s}=14$ TeV (see \autoref{fig:Pair}).   For the $bbbb$ final states, the slightly lower signal efficiencies as seen in  \autoref{tab:pair_sg_rate} are due to the stringent cut $p_T^b>400$ GeV, compared to the $tttt$ and $ttbb$ final states.

\begin{table}[tb]
    \centering
    \begin{tabular}{|c|c|c|c|c|c|c|c|}
    \hline\hline
     $\sigma$ &  $\sqrt{s}$  & \multicolumn{2}{c|}{$t\bar t b\bar b$} & \multicolumn{2}{c|}{$t\bar t t\bar t$} & \multicolumn{2}{c|}{$b\bar b b\bar b$} \\
    \cline{3-8}
    (fb) & (TeV)& $\mu^+\mu^-$ & VBF & $\mu^+\mu^-$ & VBF & $\mu^+\mu^-$ & VBF  \\
    \hline
    \multirow{3}{*}{ $H^+H^-$} & 6  & $6.7\times 10^{-4}$ & $\lesssim10^{-13} $ & $-$ & $-$ & $-$ & $-$ \\
    & 14 & $2.3\times 10^{-3}$ & $1.1\times10^{-4}$ & $-$ & $-$ & $-$ & $-$ \\
    & 30 & $1.4\times 10^{-3}$ & $5.2\times10^{-4}$ & $-$ & $-$ & $-$ & $-$ \\
    \hline
    \multirow{3}{*}{ $HA$} & 6 & $1.4\times {10^{-3}}$ & $4.0\times10^{-8}$ & $6.1\times 10^{-5}$ & $\lesssim10^{-14}$ & $1.7\times 10^{-6}$ & $\lesssim10^{-14}$ \\
    & 14 & $1.7\times 10^{-3}$ & $1.7\times10^{-4}$ & $9.0\times 10^{-4}$ & $2.5\times 10^{-5}$ & $2.7\times 10^{-5}$ & $3.9\times10^{-6}$ \\
    & 30 & $7.9\times 10^{-4}$  & $6.8\times10^{-4}$  & $6.5\times 10^{-4}$ & $1.7\times10^{-4}$ & $ 1.4\times 10^{-5}$ & $2.7\times10^{-5}$ \\
    \hline
    \end{tabular}
    \caption{Dominant background cross sections via $\mm$  annihilation and VBF processes for the signal  channels $H^+H^-$ and $HA$ with $m_\Phi=2$ TeV and $\sqrt s=6$, 14 and 30 TeV.}
    \label{tab:pair_bg_xs}
\end{table}

\begin{table}[tbh]
    \centering
    \begin{tabular}{|c|c c|c|c|c|}
    \hline\hline
    Signal Rate & $\sqrt{s}$ (TeV) & $\sigma$ (fb) & $t\bar t b\bar b$ & $t\bar t t\bar t$ & $b\bar b b\bar b$ \\
    \hline
    \multirow{3}{*}{ $H^+H^-$} & 6 & 0.32  & 70\% & $-$ & $-$  \\
    & 14& 0.14 & 79\% & $-$ & $-$  \\
    & 30 & 0.04 & 87\% & $-$ & $-$  \\
    \hline
    \multirow{3}{*}{$HA$} & 6 & 0.13 & 69\% & 81\% & 57\% \\
    & 14 & 0.06 & 79\% & 88\% & 70\%  \\
    & 30 & 0.02 & 87\% & 92\%  & 82\% \\
    \hline
    \end{tabular}
    \caption{Signal pair production cross sections and cut efficiencies for the production channels $\mm \to  H^+H^-, HA$  for $m_\Phi=2$ TeV after acceptance cuts in \autoref{eq:basic_cuts_pair}$-$\autoref{eq:HAcuts}. For $\sqrt{s}=14$ and 30 TeV, we reconstruct the heavy Higgs bosons from fermion pairs based on the smallest opening angle between two fermions, whereas for $\sqrt{s}=6$ TeV we do the reconstruction based on the invariant mass closest to $m_\Phi$. }
    \label{tab:pair_sg_rate}
\end{table}

 There are other kinematic features that could help further purify the signal samples from the backgrounds. For instance, at high energies, the Higgs bosons are produced back-to-back $\theta_{\phi_1 \phi_2} \approx \pi$, so that two fermion pairs may form a large angle. In contrast, the background is primarily from a back-to-back fermion pair production, followed by a collinear gluon splitting to the second fermion pair. Consequently, three quarks tend to cluster close by, going against another single energetic quark.

It is important to note that there are two classes of kinematic topologies for the signal of Higgs pair production, namely, the $\mm$ annihilation at high invariant mass  %\Tao{$m_{HH}\sim \sqrt s$}
and the VBF near the Higgs pair threshold,
%\Tao{$m_{HH} \sim 2m_H$},
as seen in~\autoref{fig:Pair_distri}. When needed, the invariant mass variable of the Higgs pair system can serve as a discriminator for the production mechanisms. The kinematics of the decay products, however, would look largely the same since it is governed by the heavy Higgs mass.

%%%%%%%%%%%%%%%%%%%%%%%%%%%%%%
\subsection{Distinguishing 2HDMs}

The pair production rates for all four types of 2HDMs are the same, since those only involve gauge coupling structures, or tri-Higgs couplings, which are independent of the Yukawa coupling structures. The decay branching fractions into fermion pairs, however, are different, which are determined by their Yukawa couplings characterized by tree-level parameter $\tan\beta$, as shown in~\autoref{fig:BR}. We will focus on the leading decay channels for the non-SM Higgs bosons as in~\autoref{eq:decay}.
In~\autoref{tab:pair}, we list the leading signal channels for various 2HDMs in different regions of small, intermediate and large $\tan\beta$.  Several observations can be made:
\begin{itemize}
\item{} For low values of $\tan\beta <5$, the four models cannot be distinguished since the dominating decay channels are the same: $H/A \rightarrow t\bar{t}$, $H^\pm \rightarrow tb$.
\item{} For large values of $\tan\beta>10$,
the decay modes of $H/A \rightarrow \tau^+\tau^-$, $H^\pm \rightarrow \tau\nu$ become substantial in Type-L, which can be used to separate Type-L from the other three types of 2HDMs.
\item{}  For $\tan\beta>5$, the enhancement of the bottom Yukawa coupling with $\tan\beta$ in Type-II/F leads to the growing and even the leading of $H/A\to b\bar b$ decay branching fraction, which can be used to separate Type-II/F from the Type-I 2HDM.
\item{}Type-II and Type-F cannot be distinguished for all ranges of $\tan\beta$ based on the leading channel, since the leptonic decay mode is always sub-dominate comparing to decays into top or bottom quarks in all ranges of $\tan\beta$.    The full discrimination is only possible at $\tan\beta>10$ if the sub-leading $H^\pm\to \tau\nu$ and $H/A\to \tau^+\tau^-$ decays in Type-II can be detected, which has a
branching fraction about $10\%$.
\end{itemize}

\begin{table}[tb]
    \centering
    \begin{tabular}{|c|c|c|c|c|c|}\hline
         &  production &Type-I & Type-II & Type-F & Type-L \\ \hline
         \multirow{3}{*}{small $\tan\beta<5$}&$H^+H^-$&\multicolumn{4}{c|}
         {$t\bar b, \bar t b$} \\
          &$HA/HH/AA$&\multicolumn{4}{c|}{$t\bar t, t\bar t$} \\
          &$H^\pm H/A$&\multicolumn{4}{c|}{$tb, t\bar t$} \\  \hline
          \multirow{4}{*}{intermediate $\tan\beta$}&$H^+H^-$&\multicolumn{3}{c|}{$t\bar b, \bar t b$}&$tb, \tau\nu_\tau$ \\  \cline{3-5}
         &$HA/HH/AA$&$t\bar t, t\bar t$&\multicolumn{2}{c|}{$t\bar t, b\bar b$}&$t\bar t,\tau^+\tau^-$\\
          &$H^\pm H/A$&$tb, t\bar t$&\multicolumn{2}{c|}{$tb, t\bar t$;\  $tb,b\bar b$}&$tb, t\bar t$;\ $tb,\tau^+\tau^-$; \\
          & & &  \multicolumn{2}{c|}{}& $\tau\nu_\tau, t\bar t$;\ \  $\tau\nu_\tau, \tau^+\tau^-$   \\ \hline
         \multirow{3}{*}{large $\tan\beta>10$}&$H^+H^-$& {$t\bar b, \bar t b$}& $tb, t b(\tau\nu_\tau)$&{$t\bar b, \bar t b$}&$\tau^+\nu_\tau, \tau^-  \nu_\tau$\\
         &$HA/HH/AA$&$t\bar t, t\bar t$&$b\bar{b},b\bar{b}(\tau^+\tau^-)$& {$b\bar b, b\bar b$}&$\tau^+\tau^-, \tau^+ \tau^-$\\
         &$H^\pm H/A$&$tb, t\bar t$&$tb(\tau\nu_\tau),b\bar b(\tau^+\tau^-)$& $tb,b\bar b$&$\tau^\pm  \nu_\tau,\tau^+ \tau^- $\\ \hline
    \end{tabular}
    \caption{Leading signal channels of Higgs pair production for various 2HDMs in different regions of small, intermediate and large $\tan\beta$. Channels in the parenthesis are the sub-leading channels. }
    \label{tab:pair}
\end{table}

\subsection{Muon collider reach}
\label{sec:muoncolliderreach}

 \begin{figure}[!b]
    \centering
    \includegraphics[width=\textwidth]{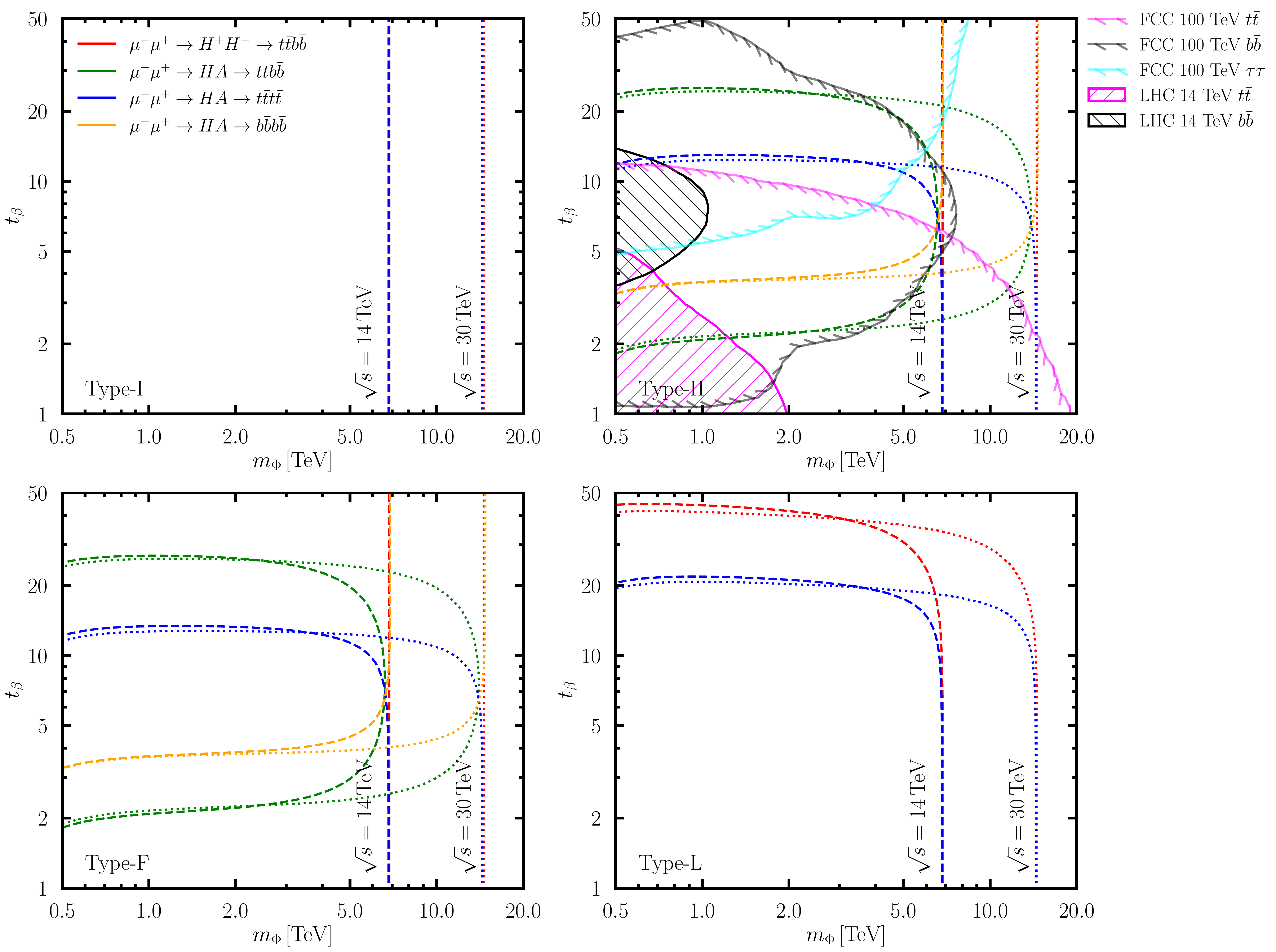}
    \caption{95\% C.L.   exclusion contour at muon collider with center of mass energy $\sqrt{s}=14$ (dash curves), 30 (dotted curves) TeV for different types of 2HDM from pair production channels with annihilation contribution only. For the Type-II 2HDM, the 95\% C.L. exclusion limits from the HL-LHC with 3 ab$^{-1}$ as well as the 100 TeV $pp$ collider with 30 ab$^{-1}$ are also shown (taken from  Ref.~\cite{Craig:2016ygr}).    }
        \label{fig:95reach}
\end{figure}

% \revision{}{This subsection is new.  Please upload the plots in overleaf.}

In~\autoref{fig:95reach}, we show the 95\% C.L. reach with the pair production process of $H^+H^-$ and $HA$ for various quark final states involving top and bottom quarks at muon collider with center of mass energy $\sqrt{s}=14$ (dash curves), 30 (dotted curves) TeV for different types of 2HDM, including the annihilation contribution only.  Note that the annihilation production channels are the dominant production modes, except for the case of large $m_\Phi$ at very high energies.
% large $\sqrt{s}$ and
Adding other channels will increase the reach, in particular, beyond the pair production mass threshold of $\sqrt{s}/2$.

\autoref{fig:95reach} shows that  reaches  in mass up to the production mass threshold of $m_\Phi \sim \sqrt{s}/2$ are possible when channels with different final states are combined.  The $\tan\beta$ dependence for different types can be understood based on the branching fraction behavior as shown in~\autoref{fig:BR}. For Type-I, the reach is independent of $\tan\beta$ since all the Yukawa couplings are modified in the same way.   Only $\mu^+\mu^-\rightarrow H^+H^-\rightarrow t\bar{t}b\bar{b}$ (red curves) and $\mu^+\mu^- \rightarrow HA \rightarrow t\bar{t}t\bar{t}$ (blue curves) are effective since $H/A \rightarrow b\bar{b}$ is suppressed comparing to the dominant $t\bar{t}$ channel.
For Type-II and Type-F, reach of $\mu^+\mu^-\rightarrow H^+H^-\rightarrow t\bar{t}b\bar{b}$ is almost independent of $\tan\beta$ since either the top or the bottom Yukawa coupling is enhanced, resulting in an almost 100\% decay branching fraction of $H^\pm \rightarrow tb$ for all values of $\tan\beta$.   $\mu^+\mu^- \rightarrow HA \rightarrow t\bar{t}t\bar{t}$ and $b\bar{b}b\bar{b}$ (orange curves) dominate in the small and large  $\tan\beta$ region, respectively, while $\mu^+\mu^- \rightarrow HA \rightarrow t\bar{t}b\bar{b}$ (green curves) only contribute in the intermediate $\tan\beta$.  For the Type-L, the reach is reduced  at large  $\tan\beta$ region given the suppressed decay branching fractions into quark final sates.   However, including the $\tau$ final state could compensate the reach significantly.

Also shown in the top-right panel of~\autoref{fig:95reach}  are the projection of 95\% C.L. exclusion reach at the HL-LHC  and 100 TeV $pp$ collider of the Type-II 2HDM,  taken from Ref.~\cite{Craig:2016ygr}.  A 14 TeV muon collide is comparable in reach to a 100 TeV $pp$ collider, except for the small $\tan\beta$, in which a 30 TeV  muon collide is comparable.

Note that the reach we obtained are based on the luminosity assumption of~\autoref{eq:luminosity}, with simple event counting and no systematic error included.  The reach scales like $\sqrt{\cal L}$ and a more thorough estimation of the muon collider reach with detailed collide simulations and systematic errors is left for future work.

%%%%%%%%%%%%%%%%%%%%%%%%%%%%%%%%%
%%%%%%%%%%%%%%%%%%%%%%%%%%%%%%%%
\section{Higgs boson associated production with a pair of heavy fermions}
\label{sec:fermion}

%%%%%%%%%%%%%%%%%%%%%%%%%%%%%%
\subsection{Production cross sections}

\begin{figure}[!tbp]
    \centering
    \includegraphics[width=0.3\textwidth]{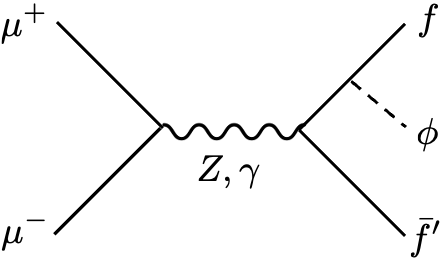}
    \caption{Representative Feynman diagram for the annihilation process: $\mm\to f\bar{f}^\prime \phi$.}
    \label{fig:Feyn_ffffs_ann}
\end{figure}

Heavy Higgs bosons can also be abundantly produced in association with a pair of heavy fermions at a muon collider. The production modes in \autoref{eq:assofermion_ann} through $\mu^+\mu^-$ annihilation are accomplished through the intermediate $\gamma^*/Z^*$ splitting into a pair of fermions, followed by the radiation of a heavy Higgs boson:
\begin{equation}
\begin{aligned}
    \mu^+\mu^- &\to b \bar b H/A,\ t \bar t H/A,\  t b H^\pm,  \\
  &\to  \tau^+ \tau^- H/A,\
  \tau^\pm  \nu_\tau H^\mp.
\end{aligned}
\label{eq:assofermion_ann}
\end{equation}
A representative Feynman diagram of the dominant contributions is shown in~\autoref{fig:Feyn_ffffs_ann}.
  The calculation is performed with tree-level diagrams. However, we include the large higher-order effects for the running of the Yukawa couplings ($Y_{u,d,e}$ in~\autoref{eq:yukawa})
   to the corresponding scale $\mu=m_\Phi$ by solving the renormalization Group Equations (RGEs) \cite{Das:2000uk}. All the input parameters listed in Sec.~\ref{sec:model} as well as the quark/lepton masses for the RGEs are given at $\mu=m_Z$~\cite{Bora:2012tx}. For $\tan\beta=1$ at $m_Z$, the running Yukawa couplings at $m_Z$, 1 TeV and 2 TeV are listed in~\autoref{tab:running_yukawa}. Effectively, compared with results using parameters at a fixed scale $m_Z$, we will have a suppression about 10\% and $14\%-30\%$ for top and bottom quark processes, respectively.

  \begin{table}[]
     \centering
     \begin{tabular}{|c|c|c|c|c|c|c|c|c|c|}
     \hline
         \multirow{2}{*}{$\tan\beta(m_Z)=1$}&\multicolumn{3}{c|}{$Y_t(\mu)$} & \multicolumn{3}{c|}{$Y_b(\mu)\,(10^{-2})$} & \multicolumn{3}{c|}{$Y_\tau(\mu)\,(10^{-2})$}  \\
         \cline{2-10}
         & $m_Z$ & 1 TeV & 2 TeV & $m_Z$ & 1 TeV & 2 TeV & $m_Z$ & 1 TeV & 2 TeV\\
         \hline
         Type-I & \multirow{4}{*}{1.40} & \multirow{4}{*}{1.33} & \multirow{4}{*}{1.32} & \multirow{4}{*}{2.29} & 2.13 & 2.10 & \multirow{4}{*}{1.42} & 1.52 & 1.54\\
         Type-II & & & & & 1.96 & 1.88 &  & 1.39 & 1.39\\
         Type-L & & & & & 2.13 & 2.10 & & 1.39 & 1.39\\
         Type-F & & & & & 1.96 &1.88 & & 1.52 & 1.54\\
         \hline
     \end{tabular}
     \caption{Running Yukawa couplings at corresponding scales for $\tan\beta(m_Z)=1$.  }
     \label{tab:running_yukawa}
 \end{table}

\begin{figure}[tb]
\centering
\includegraphics[width=\textwidth]{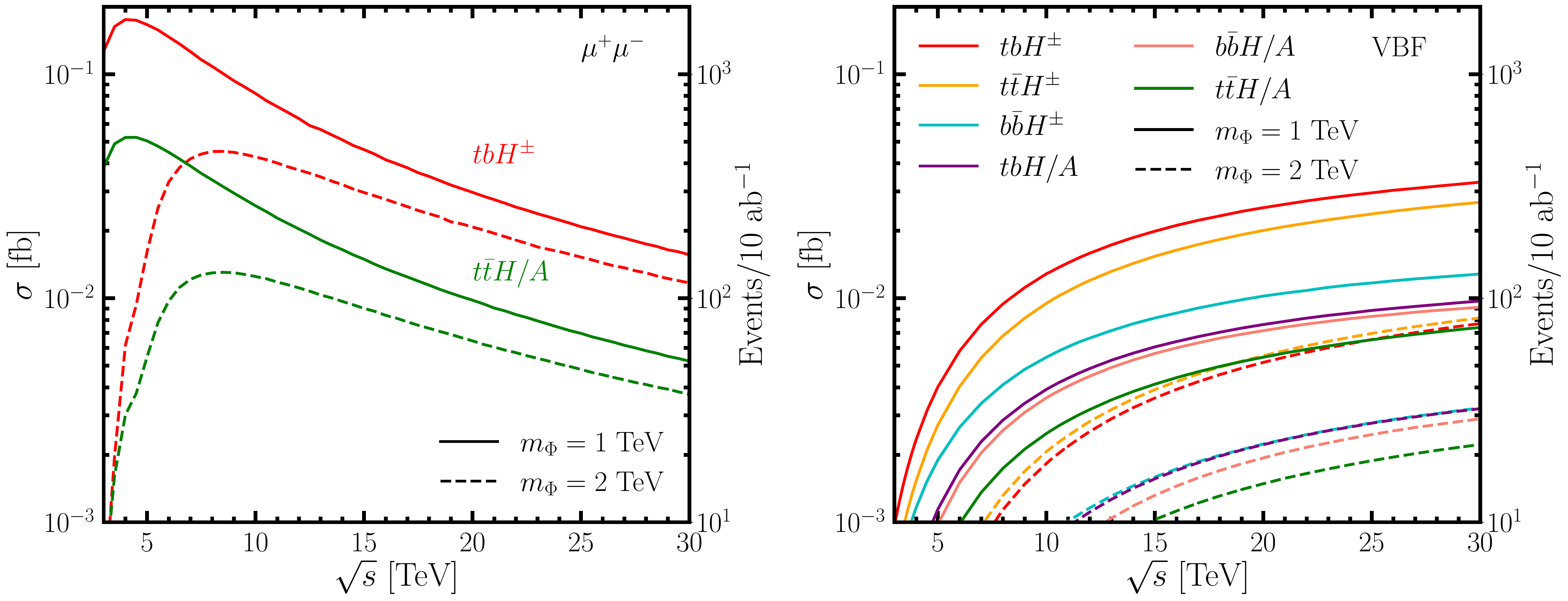}
\caption{Cross sections versus the  c.m.~energy $\sqrt{s}$ for a single heavy Higgs production associated with a pair of fermions, left panel for $\mu^+\mu^-$ annihilation, and right panel for VBF.  Acceptance cuts  of $p_{T}^f>50$ GeV and  $10^\circ<\theta_f<170^\circ$ are imposed on all outgoing fermions. A veto cut of $0.8m_\Phi<m_{f f^\prime}<1.2m_\Phi$ is applied to the associated fermions to remove contributions from resonant Higgs decays.   The vertical axis on the right shows the corresponding event yields for a 10 ab$^{-1}$ integrated luminosity.   }
\label{fig:com_ffH}
\end{figure}

  To simulate the detector acceptance in our partonic-level calculations, we first apply the simple cuts on the final state fermions
\begin{equation}
p_T^f>50 {\rm\ GeV\ and}\ 10^\circ<\theta_f<170^\circ .
\label{eq:ffH_cut1}
\end{equation}
  A veto cut of $0.8m_\Phi<m_{f f^\prime}<1.2m_\Phi$ is applied to the associated fermions to remove contributions from resonant Higgs decays.
  The signal cross sections are shown in the left panel of \autoref{fig:com_ffH}.
 We choose $\tanb=1$ so that the results are the same for four different types of 2HDMs and the Yukawa couplings of the non-SM Higgs bosons to fermions are the same as those of the SM Higgs.  Therefore, heavy quark associated productions are orders of magnitude larger than the light quark and lepton associated productions.  The dominant $tbH^\pm$ production can reach a cross section of 0.2 fb $-$ 0.02 fb for $\sqrt{s}$ between 3 $-$ 30 TeV for $m_{H^\pm}$=1 TeV.  $t\bar{t}H$ production cross section is about factor of 3 smaller.  $b\bar{b}H$ and $\tau^+\tau^- H$ cross sections are further reduced by a factor of $(m_{b,\tau}/m_t)^2$.   However, this hierarchical structure could be altered by the choice of  $\tan\beta$ in different types, which will be discussed later in Sec.~\ref{sec:dis_2hdm_ffs}.
 We see the advantage of the accessible events below the pair production threshold of $\sqrt{s}<2m_\Phi$. The production rates well above the threshold are smaller than that of the pair production via the $\mm$ annihilation in Sec.~\ref{sec:pair} by a factor of few due to the 3-body kinematics.
 The mass dependence of the single production cross sections is also stronger comparing to that of the pair production processes via annihilation, as shown in the solid curves for $m_\Phi=1$ TeV and dashed curves for $m_\Phi=2$ TeV in \autoref{fig:com_ffH}, comparing to that in the left panel of \autoref{fig:Pair}.

\begin{figure}
    \centering
    \includegraphics[width=0.8\textwidth]{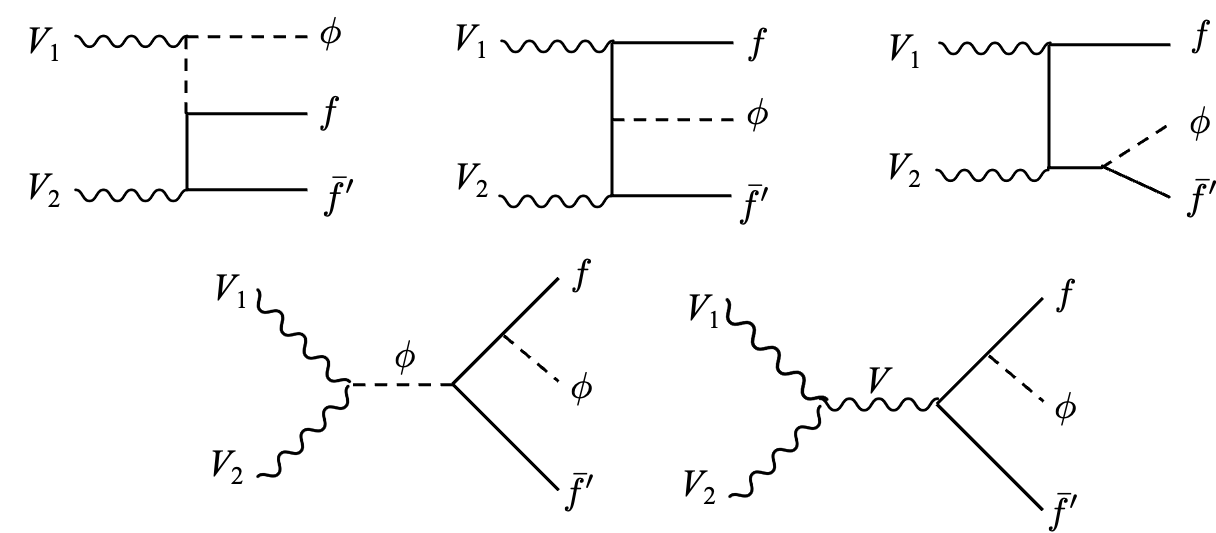}
    \caption{Representative Feynman diagrams for the VBF process $V_1V_2 \rightarrow  f\bar{f}^\prime \phi$.   }
    \label{fig:Feyn_vvffs}
\end{figure}

The fermion associated single heavy Higgs can also be produced via the VBF processes.  In addition to the charge-neutral states in \autoref{eq:assofermion_ann}, the fusions of $W^\pm\gamma/Z$ give rise to rich charged final states.
The complete set of the dominant ones are
\begin{equation}
\begin{aligned}
    \mu^+\mu^- \to VV'&\to b \bar b H/A,\ t \bar t H/A,\  t b H^\pm,\ t\bar t H^\pm, b\bar b H^\pm, tbH/A,\\
  &\to  \tau^+ \tau^- H/A,\
  \tau^\pm  \nu_\tau H^\mp,  \tau^+ \tau^- H^\pm,\tau^\pm  \nu_\tau H/A.
\end{aligned}
\label{eq:assofermion}
\end{equation}
Some representative Feynman diagrams of the dominant contributions are shown in~\autoref{fig:Feyn_vvffs}.  There also exist the pure $\nu_\tau$ associated final states such as $\nu_\tau\bar\nu_\tau H/A$ and $\nu_\tau\bar\nu_\tau H^\pm$ via VBF processes. However, given the absence of neutrino Yukawa couplings, their cross sections are about two orders of magnitude smaller than the corresponding production with $\tau$.
The simple acceptance cuts as in \autoref{eq:ffH_cut1} are again applied, which help to regularize the singularities of the outgoing fermions in the forward regions.
The cross sections for the quark associated production as a function of the c.m.~energy $\sqrt{s}$ for $\tan\beta=1$ are shown in the right panel of \autoref{fig:com_ffH}.

\begin{table}[tbh]
    \centering
    \begin{tabular}{|c c|c c|c|c|}
    \hline
    \hline
    \multicolumn{2}{|c|}{\multirow{2}{*}{$\sigma$ (fb)}} & \multicolumn{2}{|c|}{$tbH^\pm$}  &  {$t\bar t H$}  & $t\bar t H^\pm$  \\
     & & $\mu^+\mu^-$ & VBF & $\mu^+\mu^-$   & VBF \\
     \hline
     \hline
    \multicolumn{6}{|c|}{$\sqrt{s}=6$ (TeV)}\\
    \hline
    \multirow{2}{*}{$m_\Phi$ } & 1 TeV & $0.15$ & $5.3\times 10^{-3}$ & $4.5\times 10^{-2}$ &   $3.7\times 10^{-3}$ \\
     & 2 TeV & $3.3\times 10^{-2}$ & $4.7\times 10^{-4}$ & $9.7\times 10^{-3}$ &  $5.7\times 10^{-4}$ \\
    \hline
    \hline
    \multicolumn{6}{|c|}{$\sqrt{s}=14$ (TeV)}\\
    \hline
    \multirow{2}{*}{$m_\Phi$ } & 1 TeV & $5.1\times 10^{-2}$ & $1.7\times10^{-2}$ & $1.6\times 10^{-2}$   & $1.3\times10^{-2}$ \\
     & 2 TeV & $3.2\times 10^{-2}$ & $2.9\times10^{-3}$ & $9.7\times 10^{-3}$ &   $3.2\times 10^{-3}$  \\
    \hline
    \hline
    \multicolumn{6}{|c|}{$\sqrt{s}=30$ (TeV)}\\
    \hline
    \multirow{2}{*}{$m_\Phi$ } & 1 TeV & $1.6\times 10^{-2}$ & $3.0\times10^{-2}$ & $5.2\times 10^{-3}$   & $2.4\times10^{-2}$ \\
     & 2 TeV & $1.2\times 10^{-2}$ & $6.9\times10^{-3}$ & $3.7\times 10^{-3}$ &   $7.3\times10^{-3}$ \\
    \hline
    \end{tabular}
    \caption{Summary of the   leading fermion associated Higgs production cross sections via $\mu^+\mu^-$ annihilation and VBF for $m_\Phi=$1, 2 TeV and $\sqrt s=6$, 14 and 30 TeV. Acceptance  and veto cuts are the same as described in the caption of~\autoref{fig:com_ffH}.
    }
    \label{tab:higgfermion}
\end{table}

While the similar hierarchical structure of production cross section is apparent, the production cross sections also manifest the rising trend with the increasing of the c.m.~energy typical to the VBF processes. Compared to the charge-neutral final states, the charged final states have comparable cross sections. This is due to the fact that the partonic luminosities of $W\gamma$ and $\gamma\gamma$ are about the same at higher energies, which dominate the production of charged and charge-neutral final states, respectively.  The VBF production cross sections also exhibit sensitive mass dependence rough as $1/m_\Phi^2$, as shown by the  solid ($m_\Phi=1$ TeV) and dashed ($m_\Phi=2$ TeV) lines. The relative size of various VBF processes, however, could vary as $m_\Phi$ varies.  For $m_\Phi=1$ TeV, $tbH^\pm$ is larger than $t\bar{t}H^\pm$, while the order is flipped for $m_\Phi=2$ TeV.  Similarly behavior occurs for $b\bar{b}H^\pm$ versus $tbH/A$.  There are several factors contributing to this, for example, the difference in the dominant $VV$ contributions in difference processes, parton luminosity differences between the $\gamma$, $V_L$ and $V_T$, and the contributions of top and bottom Yukawa couplings that enter differently in different processes, as well as the chiral suppression of the bottom quark mass.
The cross sections of the leading   production channels are summarized in \autoref{tab:higgfermion}.
% $tbH^\pm$ has the largest production cross section for both the annihilation and VBF processes.
$tbH^\pm$ has the largest production cross section for the annihilation, while both $tbH^\pm$ and $t\bar{t}H^\pm$ contribute for the VBF processes. For the neutral Higgs production, $t\bar{t}H/A$ via annihilation is important for lower c.m.~energies, while $tbH/A$, $b\bar{b}H/A$ and $t\bar{t}H/A$ via VBF  could be important for higher c.m.~energies and low scalar masses.

\begin{figure}[tbh]
\centering
\includegraphics[width=\textwidth]{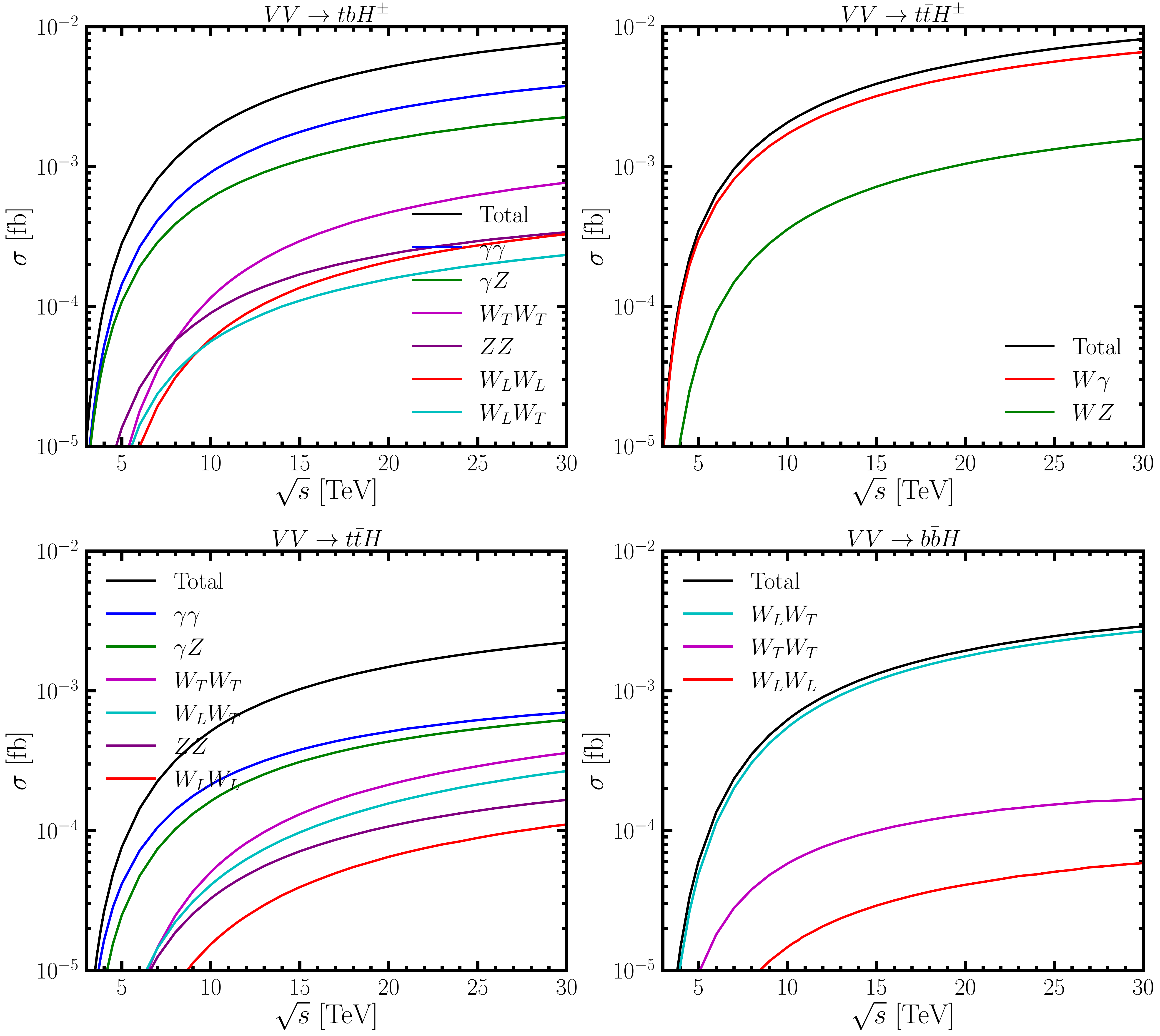}
 \caption{Cross sections versus the c.m.~energy $\sqrt s$ for individual  contributions from different VBF sub-processes, for $\tan\beta=1$ and  $m_\Phi=2$ TeV. Acceptance  and veto cuts are the same as described in the caption of~\autoref{fig:com_ffH}. }
\label{fig:fermionHiggs_ind}
\end{figure}

\begin{figure}[tbh]
\centering \includegraphics[width=0.9\textwidth]{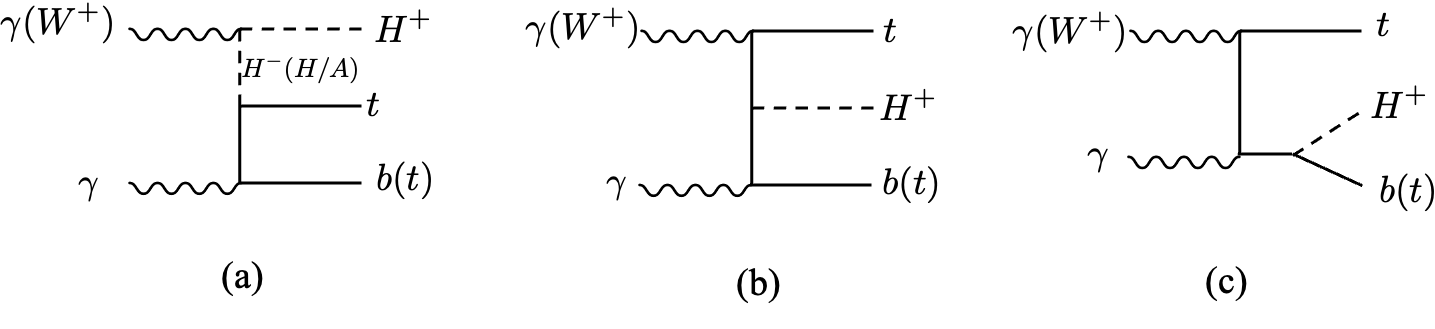}
\caption{Representative Feynman diagrams of the fermion associated production $tbH^\pm$ and $t\bar{t} H^\pm$ through VBF. }
\label{fig:feynm}
\end{figure}

One of the advantages for adopting the EW PDF approach is to appreciate the underlying contributions from the individual sub-processes.
In \autoref{fig:fermionHiggs_ind}, we plot the contributions from individual VBF processes to the charge-neutral final state process $tbH^\pm$, $t\bar{t}H$ and $b\bar{b}H$, and charged final state  process $t\bar{t}H^\pm$  with the benchmark heavy Higgs mass $m_\Phi=2$ TeV and $\tan\beta=1$.  By comparing the top two plots, we find that the production of $t\bar t H^\pm$ through $W\gamma$ fusion is slightly larger than $tbH^\pm$ through $\gamma\gamma$ fusion. That is because the  contributions from diagram (a) in \autoref{fig:feynm} is twice as large for $W\gamma$ fusion in $t\bar t H^\pm$ production with $H/A$ in the internal line, comparing to $\gamma\gamma$ fusion in $tb H^\pm$ production with $H^\pm$ in the internal line.  $\gamma\gamma$ fusion contribution in $t\bar{t}H$ is much smaller since  contribution from diagram (a) is absent.  Furthermore, $\gamma\gamma$ contribution to the $b\bar b H$ production (bottom right panel) is even smaller since $H$ couples to the internal line through the small bottom Yukawa coupling.
The $WW$ fusion similar to diagram (b) in \autoref{fig:feynm}, however, is dominant since $H$ couples to the internal line through large top Yukawa coupling. Given that the longitudinal component of $W$ couples to heavy quarks through Yukawa coupling, and the largeness of $W_T$ parton luminosity, the contribution through $W_LW_T$ fusion dominates over the other fusion processes.

\begin{figure}[tb]
\centering
\includegraphics[width=\textwidth]{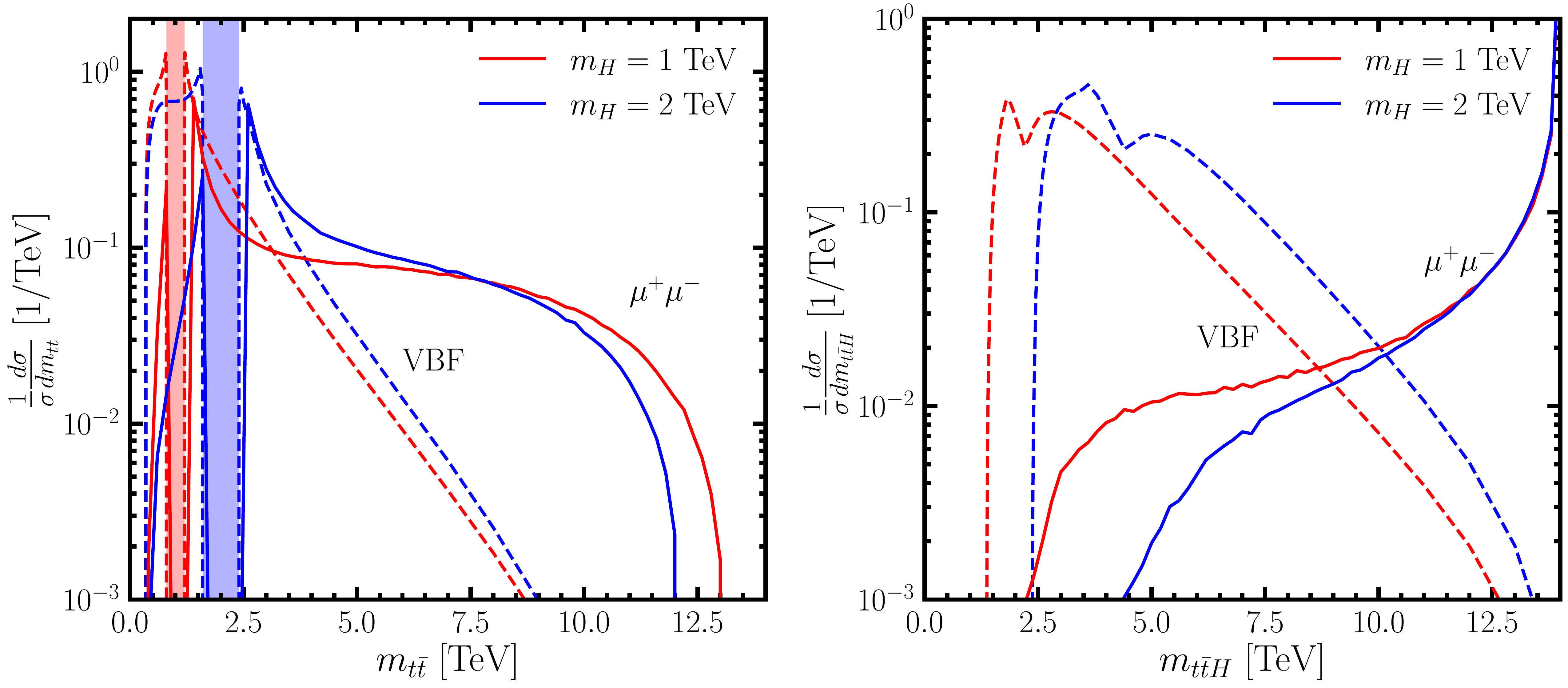}
\caption{Normalized invariant mass distributions for $\mu^+\mu^-\to t\bar t H$ at $\sqrt{s}=14$ TeV, left panel for $m_{t\bar t}$ and right panel for $m_{t\bar t H}$, for $\mu^+\mu^-$-annihilation channels (solid) and VBF channels (dashed), respectively. Red and blue curves are for $m_H=1$ and 2 TeV, respectively.
}
\label{fig:dcross_mtt_ttH}
\end{figure}

 \autoref{fig:dcross_mtt_ttH} shows the normalized differential cross sections of $m_{t\bar{t}}$ (left panel) and $m_{t\bar{t}H}$ (right panel) for $\mu^+\mu^-\to t\bar t H$ process   at $\sqrt{s}=14$ TeV. Red and blue curves are for $m_{H}=1$ and 2 TeV, respectively. The two shaded bands at $[0.8m_H,1.2m_H]$ of consistent colors indicate the regions where the two associated tops originating from resonant heavy Higgs decay, which are excluded from the production cross section by the $m_{t\bar{t}}$ cut.  We use dashed and solid lines to indication the production of VBF and $\mu^+\mu^-$-annihilation channels.  The $m_{t\bar{t}}$ distributions of VBF channels mostly peak at small invariant mass regimes while those of $\mu^+\mu^-$-annihilation channels have flat distribution till threshold of $\sqrt{s}-m_H$. The $m_{t\bar{t}H}$ distributions (right panel) exhibit similar feature as the pair production process:  the distributions are peaked at $\sqrt{s}$ for annihilation process,  while peaked at the production threshold $m_H+2m_t$ for the VBF process.

%%%%%%%%%%%%%%%%%%%%%%%%%%%%%%
\subsection{Signals and backgrounds}

Following the discussions of the signal construction and background suppression as in Sec.~\ref{sec:SB}, we present the scheme to identify the Higgs boson signal from the associated production with a pair of heavy fermions, again illustrated for $m_\Phi=2$ TeV and $\sqrt{s}=14$ TeV.
We first note that the four-fermion background processes are of the same origin as in Sec.~\ref{sec:SB}. As for the signal, there is only one heavy Higgs boson in the events.
As an illustration, we focus on the two leading production channels $tbH^\pm$ and $t\bar t H$.
We first impose the basic acceptance cuts of $p_T$ and $\theta$ cuts in \autoref{eq:ffH_cut1} on all the final state fermions,
and then propose the appropriate cuts to suppress the irreducible backgrounds for the dominate decay modes.

 %%%%%%%%%%%%%%%%%%%%%%%%%%%%%
    \subsubsection{$tbH^\pm\to t \bar b\ \bar t b$}

       \begin{figure}[tbh]
\centering
\includegraphics[width=0.49\textwidth]{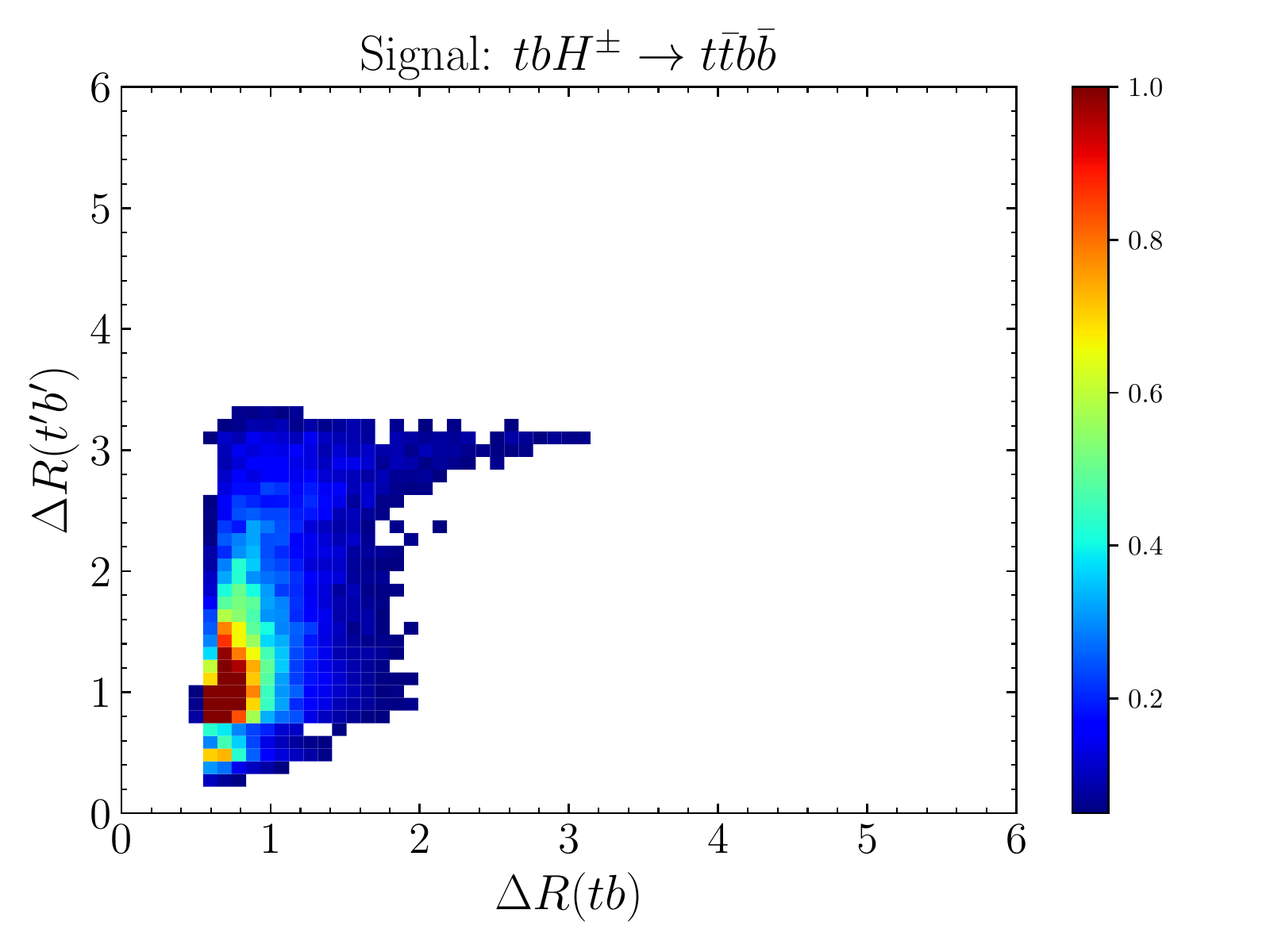}
\includegraphics[width=0.49\textwidth]{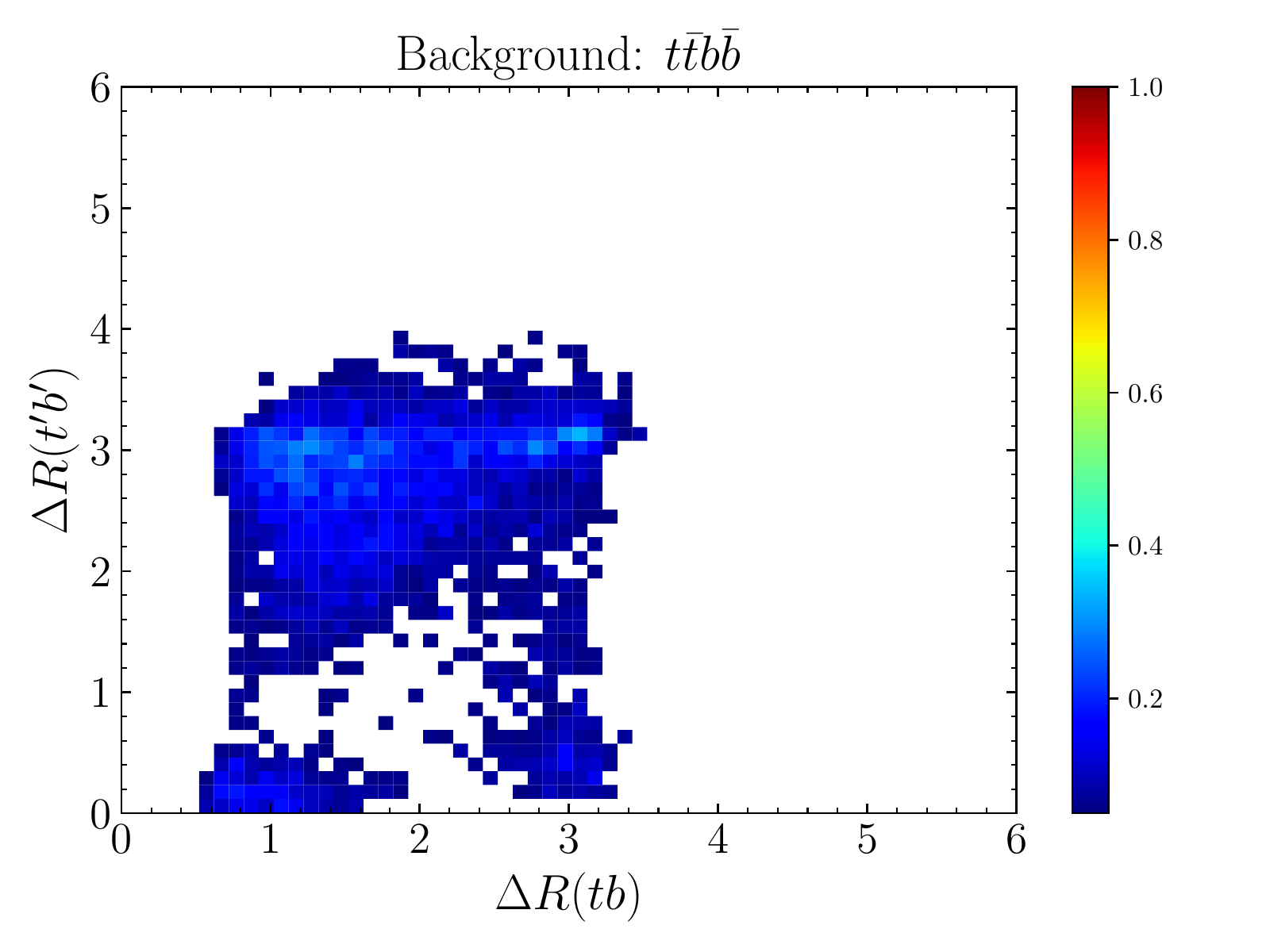}
\includegraphics[width=0.49\textwidth]{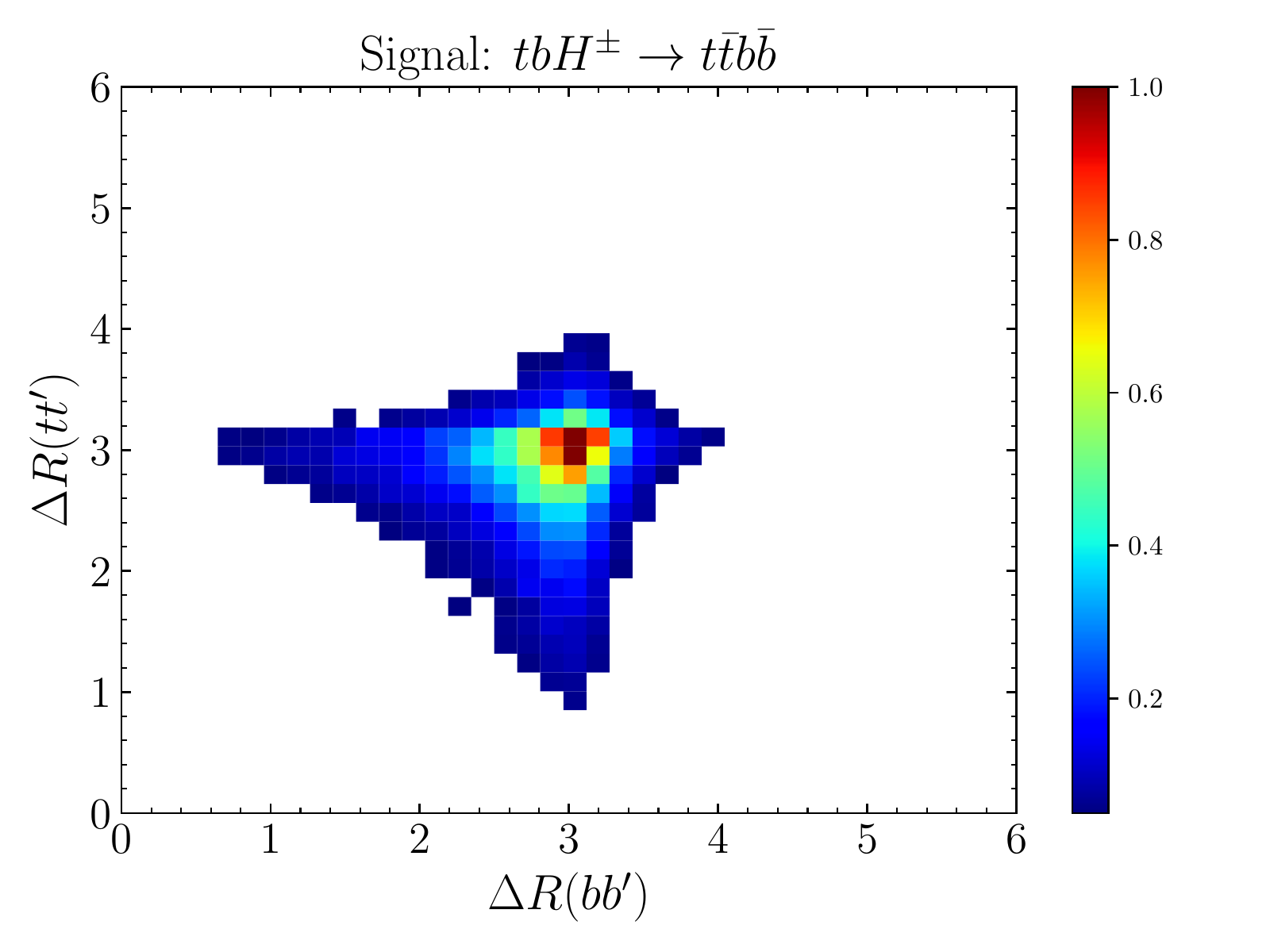}
\includegraphics[width=0.49\textwidth]{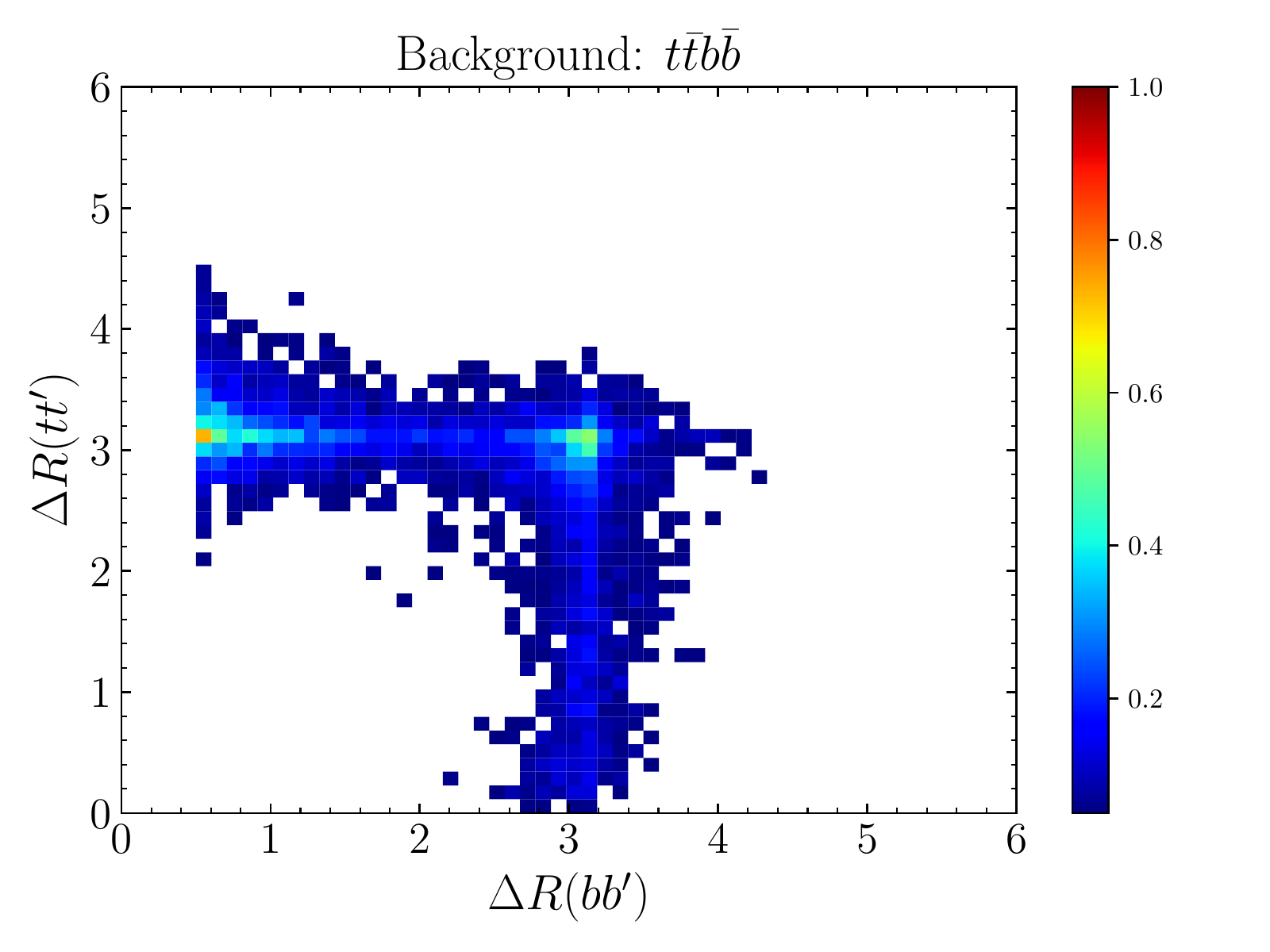}
\caption{Event distributions of the annihilation process for the $\mm \to tbH^\pm(\to tb)$ signal (two left panels) and the $\mm \to t\bar t b\bar b$  background (two right panels) in the planes of $\Delta R(tb)$-$\Delta R(t'b')$ (top panels) and  $\Delta R(tt)$-$\Delta R(bb)$ (bottom panels), with $m_\Phi=2$ TeV and $\sqrt{s}=14$ TeV. The invariant mass cut $m_{tb}>0.9m_{H^\pm}$ is imposed to the background.  }
\label{fig:tbhpm3}
\end{figure}

   We first consider the signal production via the $\mu^+\mu^-$ annihilation.
    For the signal reconstruction,  we find the pair of $tb$ which gives the closest invariant mass to the hypothetical  $m_{H^\pm}$.
    The other pair of $tb$ unlikely to be from the Higgs decay is denoted by $t'b'$. In the upper left panel of \autoref{fig:tbhpm3}, we present the distribution for the scatter events of $\mu^+\mu^-\to t'b' H^\pm(\to tb)$ in the plane of $\Delta R(tb)$-$\Delta R(t'b')$ for the signal (left panel) and the background (right panel).
     For the signal process, while $tb$
     tends to form a small angle due to the  energy boost $\theta_{tb}\sim  m_H/E_H$, the angle between $t'b'$ varies over a wide range. The most populous region conveys the picture of a heavy Higgs flying in one direction and the associated fermions go oppositely to balance the momenta. The upper right panel of \autoref{fig:tbhpm3} shows the scatter plot distribution for the $t\bar t b\bar b$ background after a cut $m_{tb}>0.9m_H$. Comparing it with the left panel, we find that they exhibit very distinct kinematic features.

\begin{table}[tb]
    \centering
    \begin{tabular}{|c|c|c|c|c|c| }
    \hline\hline
    \multirow{2}{*}{$\sigma$ (fb)} & \multirow{2}{*}{$\sqrt{s}$ (TeV)}& \multicolumn{2}{c|}{$t\bar t b\bar b$} & \multicolumn{2}{c|}{$t\bar t t\bar t$}  \\
    \cline{3-6}
    & & $\mu^+\mu^-$ & VBF & $\mu^+\mu^-$ & VBF   \\
    \hline
    \multirow{3}{*}{ $tbH^\pm$} & 6  & $1.8\times 10^{-3}$ & $1.2\times10^{-4}$ & $-$ & $-$    \\
    & 14 & $1.7\times 10^{-3}$ & $8.8\times10^{-4}$ & $-$ & $-$   \\
    & 30 & $7.7\times 10^{-4}$ & $2.0\times10^{-3}$ & $-$ & $-$  \\
    \hline
    \multirow{3}{*}{ $t\bar tH$} & 6 & $1.1\times 10^{-3}$ & $1.4\times10^{-4}$ & $9.5\times 10^{-4}$ & $1.1\times10^{-3}$ \\
    & 14 & $7.7\times 10^{-4}$ & $9.2\times10^{-4}$ & $8.6\times 10^{-4}$ & $4.6\times10^{-3}$   \\
    & 30 & $3.0\times 10^{-4}$  & $2.0\times 10^{-3}$  & $3.5\times 10^{-4}$ & $8.7\times10^{-3}$   \\
    \hline
    \end{tabular}
    \caption{Dominant background cross sections via $\mm$  annihilation and VBF processes for the signal  channels $tb H^\pm$ and $t \bar t H$ with $m_\Phi=2$ TeV and $\sqrt s=6$, 14 and 30 TeV.
    }
    \label{tab:bg_XS_ffH}
\end{table}

\begin{table}[tb]
    \centering
    \begin{tabular}{|c|c|c|c|c|c|c|c|}
    \hline\hline
    \multirow{2}{*}{Signal Rate} & \multirow{2}{*}{$\sqrt{s}$ (TeV)} & \multicolumn{2}{c|}{$\sigma$ (fb)} & \multicolumn{2}{c|}{$t\bar t b\bar b$} & \multicolumn{2}{c|}{$t\bar t t\bar t$} \\
    \cline{3-8}
    & & $\mu^+\mu^-$ & VBF & $\mu^+\mu^-$ & VBF & $\mu^+\mu^-$ & VBF  \\
    \hline
    \multirow{3}{*}{ $tbH^\pm$} & 6 & $3.7\times 10^{-2}$ & $5.3\times 10^{-4}$ & 28\% & 35\% & $-$ & $-$  \\
    & 14 & $3.6\times 10^{-2}$ & $4.1\times 10^{-3}$ & 72\% & 48\% & $-$ & $-$ \\
    & 30 & $1.3\times 10^{-2}$ & $1.1\times 10^{-2}$ & 77\% & 53\% & $-$ & $-$ \\
    \hline
    \multirow{3}{*}{ $t\bar tH$} & 6 & $1.1\times 10^{-2}$ & $1.4\times 10^{-4}$ & 48\% & 75\% & 27\% & 75\%  \\
    & 14 & $1.1\times 10^{-2}$ & $9.3\times 10^{-4}$ & 72\% & 72\% & 62\% & 72\%  \\
    & 30 & $4.1\times 10^{-3}$ & $2.2\times 10^{-3}$ & 71\%  & 67\%  & 60\% & 67\% \\
    \hline
    \end{tabular}
    \caption{Signal cross sections for single Higgs  production in associated with fermions and cut efficiencies for the $ tbH^\pm$ and $t\bar{t}H$ channels via $\mu^+\mu^-$ annihilation and VBF for $m_\Phi=$2 TeV and $\sqrt s=6$, 14 and 30 TeV.}
    \label{tab:signalrate_ffH}
\end{table}

To further separate the signal and background, in the bottom two panels of \autoref{fig:tbhpm3}, we depict their distributions in the plane of $\Delta R(t t')$-$\Delta R(b b')$.    Again, we see the back-to-back feature of the two $tb$ pairs for the signal,
namely, most signal events concentrate at $\theta_{tt'}= \pi$ and $\theta_{bb'}= \pi$, because $t$ and $b$ originating from the $H^\pm$ decay tends to fly in the same direction due to the high energy boost. However background events own two concentrations in the lower right panel, the one with $bb'$ from the gluon splitting, and the one with $tb$ from $W$ decay.
Based on these distributions, we thus propose the following additional cuts to suppress the $t\bar tb\bar b$ background for the $tbH^\pm$ channel for the benchmark Higgs mass of 2 TeV:
    \begin{equation}
    \begin{aligned}
        &\ p_T^{t_1} > 100~{\rm GeV},\ \ p_T^{b_1}> m_\Phi/5,\ m_{tb}>0.9m_{H^\pm},\\
        &\Delta R(tb)<2.0,\ \Delta R(t'b')<2.8,\ \Delta R(tt')>1.5,\ \Delta R(bb')>1.0.
    \end{aligned}
    \label{eq:ttHC_ANN_cut}
    \end{equation}
    where $t_1$ and $b_1$ are the top and bottom quarks with leading $p_T$. We note that the leading $p_T$ fermions may not be from the heavy Higgs decay in this production mechanism.
    With the above cuts, we can suppressed the $t\bar t b\bar b$ background down to the level of $10^{-3}$~fb, while retaining the signal at the level of $10^{-2}$~fb. The cuts are more efficient at high c.m.~energy since they're proposed based on the specific features of the event distribution manifested at high enough energy. At $\sqrt{s}= 14$ TeV, we can achieve a signal rate up to 72\%.  However, at low c.m.~energies, the distributions shown in \autoref{fig:tbhpm3} would be less distinguishable. The optimal cut selection should be properly tuned according to the hypothetical Higgs mass and the c.m.~energy.
    The background cross sections after the cuts are given in \autoref{tab:bg_XS_ffH} and the corresponding signal rates are shown in \autoref{tab:signalrate_ffH}.

    For the VBF processes, the system is close to the production threshold and thus the final state quarks are less boosted. The angular cuts should be adjusted  accordingly.
    For the top and bottom quarks that are paired from the heavy Higgs decay, we require
    \begin{align}
        p_T^t > 100\,{\rm GeV},\,p_T^b > m_\Phi/5,\,\Delta R(tb) < 3.0,\, m_{tb} > 0.9 m_{H^\pm}
        \label{eq:ttHC_VBF_cut}
    \end{align}
    For the other pair of top and bottom quarks, we require
    \begin{align}
        p_T^{t'} > 50\,{\rm GeV},\,p_T^{b'} > 50\,{\rm GeV},\,\Delta R(t'b') > 0.5
    \end{align}
    Then for the two tops and two bottoms, we require:
    \begin{align}
        \Delta R(t t')>1.0,\,\Delta R(b b')>1.0.
    \end{align}
We again achieve efficient signal-background separation, as presented in \autoref{tab:signalrate_ffH} and \autoref{tab:bg_XS_ffH} for the signal and background, respectively.

    %%%%%%%%%%%%%%%%%%%%%%%%%%%%%%%%%%%%%
    \subsubsection{$t\bar t H\to t\bar t b\bar b$}
    We again first consider the annihilation production.
    One advantage of this decay mode is that the heavy Higgs reconstruction is free of the combinatorial problem. Based on a similar analysis as in the previous section, in additional to the basic acceptance in \autoref{eq:ffH_cut1},
    we implement the following cuts to suppress the  background
    \begin{equation}
    \begin{aligned}
        &\ p_T^{t_1} > 100~{\rm GeV},\ \ p_T^{b_1}> m_\Phi/5,\ m_{bb}>0.9m_{H},\\
        &\Delta R(bb)<2.5,\ \Delta R(tt)<3.0,\ \Delta R(t_2b_2)>1.0,
    \end{aligned}
    \end{equation}
    where $t_2$ and $b_2$ are top and bottom quarks sub-leading in $p_T$.

    For the VBF process, the cuts are accordingly adjusted as
    \begin{align}
        p_T^{b_{1,2}} > 400\,{\rm GeV},\,\Delta R(t_1t_2) > 1.5,\,m_{bb} > 0.9m_H .
    \end{align}
    The achieved signal and background separation, as presented in \autoref{tab:signalrate_ffH} and \autoref{tab:bg_XS_ffH} for the signal and background, respectively.

    %%%%%%%%%%%%%%%%%%%%%%%%%%%%%%%%%%%
    \subsubsection{$t\bar t H\to t\bar t t\bar t$}
    Tops sorted by $p_T$ from high to low are labeled by $t_1, t_2,t_3,t_4$. The top pair chosen to reconstruct heavy Higgs are denoted by $tt$ while the other two tops are denoted by $t't'$. We reconstruct heavy Higgs based on invariant mass close to the hypothetical $m_\Phi$ in the analyses. For the $\mm$  annihilation production, the cuts we propose to suppress the $t\bar t t\bar t$ background are
    \begin{equation}
    \begin{aligned}
        &\ p_T^{t_{1,2}} > 100~{\rm GeV},\ m_{tt}>0.9m_{H},\\
        &\Delta R(tt)<2.0,\ \Delta R(t't')<2.8,\ \Delta R(t_3t_4)>1.0 .
    \end{aligned}
    \end{equation}
    With those cuts, we find that the overall event efficiencies are slightly lower than the $t\bar tb\bar b$ final states due to the combinatorics,  ranging from  62\% compared to 72\% at $\sqrt{s}=14$ TeV and $60\%$ compared to 71\% at $\sqrt{s}=30$ TeV.

    For the VBF production, the cuts are slightly different,
    \begin{align}
        p_T^t > 100\,{\rm GeV},\,m_{tt}>0.9m_H,\,\Delta R(t't')>2.0.
    \end{align}
  The achieved signal and background separation, as presented in \autoref{tab:signalrate_ffH} and \autoref{tab:bg_XS_ffH} for the signal and background, respectively.

 In summary, we have demonstrated that it is quite  conceivable to archive very desirable signal identification over the SM backgrounds for the associated heavy Higgs production for both direct $\mm$ annihilation and the VBF channels.

%%%%%%%%%%%%%%%%%%%%%%%%%%%%%%
\subsection{Distinguishing 2HDMs}
\label{sec:dis_2hdm_ffs}

\begin{figure}[tb]
\centering
\includegraphics[width=\textwidth]{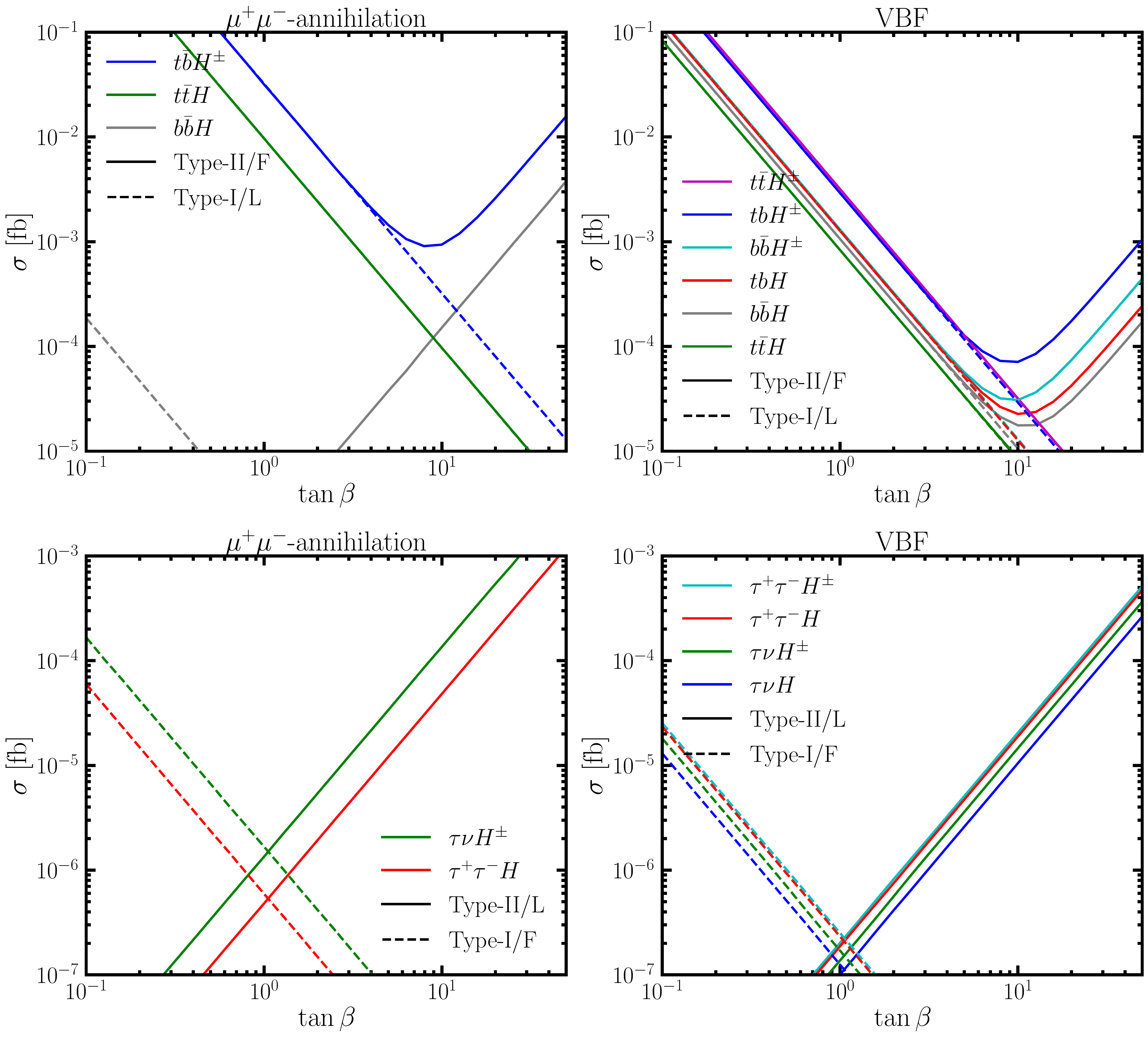}
\caption{The $\tan\beta$ dependence of the leading Higgs production cross sections in association with a pair of heavy fermions at $\sqrt{s}=14$ TeV, $m_\Phi=2$ TeV and $\cos(\beta-\alpha)=0$. The left panels show the annihilation production and the right panels show the VBF production. }
\label{fig:quark_tanb}
\end{figure}

The production of heavy Higgs bosons in association with fermions is achieved through Yukawa couplings, thus the cross sections are sensitive to $\tan\beta$, which behave differently for different types of 2HDMs.   In \autoref{fig:quark_tanb}, we demonstrate the $\tan\beta$ dependence of quark-associated production channels in the top two panels and $\tau$ associated production channels in the bottom two panels. The left panels are for the production through $\mu^+\mu^-$ annihilation and the right panels are the production through VBF. As a benchmark point, we choose degenerate heavy Higgs mass $m_\Phi=2$ TeV ($\Phi=H,A,H^\pm$) and the collider c.m.~energy $\sqrt{s}=14$ TeV.

The $\tan\beta$ dependence for the production through $\mu^+\mu^-$ annihilation is directly related to the Yukawa couplings in~\autoref{eq:normcoupling}. For the quark associated production, they're uniformly proportional to $1/\tan^2\beta$ in Type-I/L.   In Type-II/F, the $tt$ and  $bb$-associated production is proportional to $(Y_t/\tan\beta)^2$ and $(Y_d\tan\beta)^2$, respectively.  The charged Higgs production $tbH^\pm$ scales with $(Y_b\tan\beta)^2+(Y_t/\tan\beta)^2$.

For the production through VBF, diverse production diagrams in \autoref{fig:Feyn_vvffs} make the $\tan\beta$ dependence more complicated, but the overall behaviours can be explained by the large contributions from diagram (a) in~\autoref{fig:feynm}, unless it is severely suppressed   by the small Yukawa coupling, or the absence of the $V\phi_1\phi_2$ coupling.

For the $\tau$ associated production, the cross sections scale as $1/\tan^2\beta$ in Type-I/F, and $\tan^2\beta$ in Type-II/L.  They are only large enough for observation in Type-II/L at large $\tan\beta$ region.

\begin{table}[tb]
    \centering
    \begin{tabular}{|c|c|c|c|c|c|}
        \hline
         &  production &Type-I & Type-II & Type-F & Type-L \\
         \hline
         \multirow{6}{*}{small $\tan\beta<5$} & $tbH^\pm$    & \multicolumn{4}{c|}{$tb, tb$} \\
         & $t\bar{t}H^\pm$   & \multicolumn{4}{c|}{$t\bar{t}, tb$} \\
         & $b\bar{b}H^\pm$   & \multicolumn{4}{c|}{$b\bar{b}, tb$} \\
        &$t\bar{t}H/A$ & \multicolumn{4}{c|}{$t\bar t, t \bar{t}$}\\
        &$b\bar{b}H/A$ & \multicolumn{4}{c|}{$b\bar b, t \bar{t}$}\\
        &$tbH/A$     & \multicolumn{4}{c|}{$tb, t \bar{t}$}\\               \hline
          \multirow{6}{*}{intermediate $\tan\beta$} &$tbH^\pm$ &\multicolumn{3}{c|}{$tb,tb$}&$tb,tb$; $tb,\tau\nu_\tau$ \\
          \cline{3-6}
           &$t\bar{t}H^\pm$ &\multicolumn{3}{c|}{$t\bar{t},tb$}&$t\bar{t},tb$; $t\bar{t},\tau\nu_\tau$ \\
          \cline{3-6}
            &$b\bar{b}H^\pm$ &\multicolumn{3}{c|}{$b\bar{b},tb$}&$b\bar{b},tb$; $b\bar{b},\tau\nu_\tau$ \\
          \cline{3-6}
                                                    &$t\bar{t}H/A$& $t\bar{t}, t\bar{t}$&\multicolumn{2}{c|}{$t\bar{t},t\bar{t}$; $t\bar{t}, b\bar{b}$} &$t\bar{t}, t\bar{t}$; $t\bar{t},\tau^+\tau^-$ \\
         &$b\bar{b}H/A$& $b\bar{b}, t\bar{t}$&\multicolumn{2}{c|}{$b\bar{b}, t\bar{t}$; $b\bar{b}, b\bar{b}$} &$b\bar{b}, t\bar{t}$; $b\bar{b},\tau^+\tau^-$\\
         &$tbH/A$& $tb,t\bar{t}$&\multicolumn{2}{c|}{$tb,t\bar{t}$; $tb,b\bar{b}$} &$tb,t\bar{t}$; $tb,\tau^+\tau^-$ \\
        \hline
          \multirow{4}{*}{large  $\tan\beta>10$}& $tbH^\pm$ & $-$ & {$tb,tb (\tau\nu_\tau)$} & {$tb,tb$} & $-$\\
          \cline{4-5}
          & $bbH^\pm$ &$-$ & {$bb,tb (\tau\nu_\tau)$} & {$bb,tb$} & $-$\\
          \cline{4-5}
          &$b\bar{b}H/A$&$-$ & {$b\bar{b}, b\bar{b} (\tau^+\tau^-)$}&{$b\bar{b}, b\bar{b}$}& $-$\\
          \cline{4-5}
          &$t\bar{b}H/A$&$-$ & {$t\bar{b}, b\bar{b} (\tau^+\tau^-)$}&{$t\bar{b}, b\bar{b}$}& $-$\\
          \hline
\multirow{4}{*}{very large  $\tan\beta>50$}
    &$\tau\nu_\tau H^\pm$& \multicolumn{3}{c|}{$-$}& $\tau\nu_\tau, \tau\nu_\tau$\\
    &$\tau^+\tau^- H^\pm$& \multicolumn{3}{c|}{$-$}& $\tau^+\tau^-, \tau\nu_\tau$\\
    &$\tau^+\tau^- H/A$& \multicolumn{3}{c|}{$-$}& $\tau^+\tau^-, \tau^+\tau^-$\\
     &$\tau\nu_\tau H/A$& \multicolumn{3}{c|}{$-$}& $\tau \nu_\tau, \tau^+\tau^-$\\
        \hline
    \end{tabular}
    \caption{Leading signal channels of single Higgs associated production with a pair of fermions for various 2HDMs in different regions of small, intermediate and large $\tan\beta$.  Channels in the parenthesis are the sub-leading channels.
    }
    \label{tab:fermion}
\end{table}

In \autoref{tab:fermion} we summarized the leading signal channels of the Higgs associated production with fermions in four types of 2HDMs in different regimes of $\tan\beta$.  Several observations can be made:
\begin{itemize}
    \item In the small $\tan\beta<5$ region, all six production channels have sizable production cross sections.  However, it is hard to distinguish different types of 2HDMs since they all lead to the same final states.
    \item In the large $\tan\beta>10$ region, all the production channels for the Type-I are suppressed, while Type-II/F have sizable production in $tbH^\pm$, $bbH^\pm$, $bbH/A$ and $tbH/A$ channels.  Type-II and Type-F can be further separated by studying the sub-dominant decay channels of $H^\pm \rightarrow \tau \nu_\tau$ and $H/A \rightarrow \tau^+\tau^-$ in the Type-II.   Same final states of Type-II can also be obtained via $\tau\nu_\tau H^\pm$, $\tau^+\tau^- H^\pm$, $\tau^+\tau^- H/A$ and $\tau\nu_\tau H/A$ production.
    \item The intermediate range of $\tan\beta$ is the most difficult
    region for all types of 2HDMs, since top Yukawa couplings are reduced, while bottom Yukawa coupling is not big enough to compensate, resulting in a rather low signal production rate.  A rich set of final states, however, are available given the various competing decay modes of $H^\pm$ and $H/A$.
    \item At very large value of $\tan\beta>50$, the tau-associated production $\tau\nu_\tau H^\pm$, $\tau^+\tau^- H^\pm$, $\tau^+\tau^- H/A$ and $\tau\nu_\tau H/A$ would be sizable for Type-L.
\end{itemize}

%%%%%%%%%%%%%%%%%%%%%%%%%
\section{Radiative return}
\label{sec:radiative}

\begin{figure}[tbh]
\begin{center}
\includegraphics[width=0.25\textwidth]{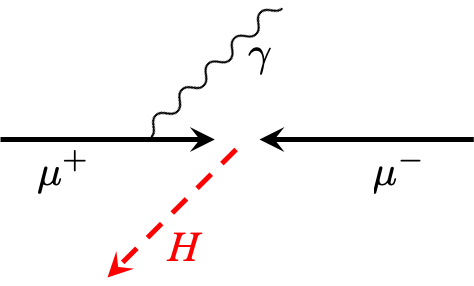}
\end{center}
\caption{Illustrative diagram for the radiative return to resonant production of a heavy Higgs boson with ISR.}
\label{fig:radiative}
\end{figure}

While the cross sections for heavy Higgs pair production are un-suppressed under the alignment limit, the cross section has a threshold cutoff at $m_\Phi\sim \sqrt{s}/2$.   The resonant production for a single heavy Higgs boson may  further extend the coverage to about
$m_\Phi\sim \sqrt{s}$, as long as the coupling strength to $\mu^+\mu^-$ is big enough. The drawback for the resonant production is that the collider energy would have to be tuned close to the mass of the heavy Higgs, which is less feasible at future muon colliders. A promising mechanism is to take advantage of the initial state radiation (ISR), so that the colliding energy is reduced to a lower value for a  resonant production, thus dubbed the ``radiative return''~\cite{Chakrabarty:2014pja}, as shown in~\autoref{fig:radiative}.

\begin{figure}[tb]
\centering
\includegraphics[width=\textwidth]{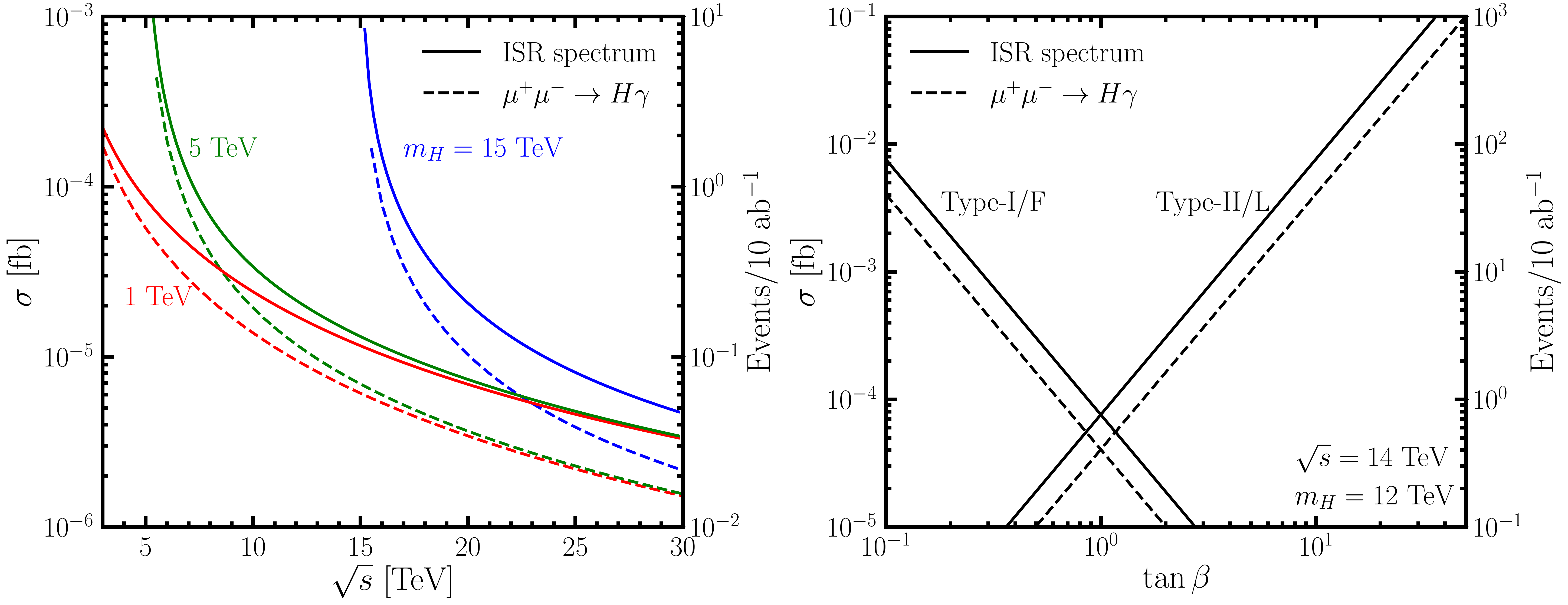}
\caption{Cross sections of  single heavy Higgs $H$ production through the radiative return. Left panel is for $m_H=1$, 2 and 15 TeV and $\tan\beta=1$ versus the c.m.~energy $\sqrt s$, with the solid curves for the convoluted ISR spectrum and the dashed curves for single photon radiation $\mu^+\mu^- \rightarrow H \gamma$ with $10^\circ<\theta_\gamma<170^\circ$.  Right panel is for the $\tan\beta$ dependence of the cross section for $\sqrt{s}=14$ TeV and $m_H=12$ TeV. The vertical axis on the right shows the corresponding event yields for a 10 ab$^{-1}$ integrated luminosity.}
\label{fig:gammaH}
\end{figure}

 This mechanism can be characterized by the process
\begin{equation}
    \mu^+\mu^-\to \gamma H,
    \label{eq:radreturn}
\end{equation}
where $\gamma$ can be a mono-photon observed in the detector, or unobserved along the beam as the collinear radiation.
We first calculate the cross section of the mono-photon process
for $m_H=1$, 5, 15 TeV at $\tan\beta=1$. $10^\circ<\theta_\gamma<170^\circ$ is imposed for the photon detection acceptance. For a single photon production, its energy is mono-chromatic $E_\gamma = (s-m_H^2)/2\sqrt s$. The results are given in the left panel of \autoref{fig:gammaH} by the dashed curves.

As a comparison, we calculate the $\mu^+\mu^-\to H$ process with ISR spectrum
\begin{equation}
    f_{\mu/\mu}(x)=\frac{\alpha}{2\pi}\frac{1+x^2}{1-x}\log\frac{s}{m_\mu^2} ,
    \label{eq:ISR_spec}
\end{equation}
where $x$ is the energy fraction carried by the muon after the ISR.  The partonic cross section is
\begin{equation}
    \hat\sigma(\mu^+\mu^-\to H)=\frac{\pi Y_\mu^2}{4}\delta(\hat s-m_H^2)=\frac{\pi Y_\mu^2}{4s}\delta(\tau - {m_H^2\over s}).
\end{equation}

To compare with process in \autoref{eq:radreturn}, we calculate the cross section to the first order of $\alpha$ by convoluting  the ISR spectrum to one muon beam,
\begin{equation}
\sigma =2\int dx_1 f_{\mu/\mu}(x_1)\hat \sigma(\tau=x_1) = \frac{\alpha Y_\mu^2}{4s} \frac{s+m_H^4/s} {s-m_H^2} \log\frac{s}{m_\mu^2}.
\end{equation}
The results are given in the left panel of \autoref{fig:gammaH} by the solid curves.  The cross section increases as the heavy Higgs mass approaches the collider c.m.~energy, closer to the $s$-channel resonant production.

The right panel of~\autoref{fig:gammaH} shows the   $\tan\beta$ dependence of the cross section for $\sqrt{s}=14$ TeV and $m_H=12$ TeV.
While the cross section at $\tan\beta=1$ is much smaller than the other production channels we considered earlier, the cross section scales like $\tan^2\beta$ in Type-II/L, which could be sizable at large $\tan\beta$. It could even be the dominant production for heavy Higgs in the large $\tan\beta$ region of Type-L, when pair production is kinematically forbidden and quark associated productions are suppressed.

%%%%%%%%%%%%%%%%%%%%%%%%%%%%%%%%%%%%%
\section{Summary}
\label{sec:conclusion}
High energy muon colliders offer new opportunities for the direct production of heavy particles.  In this paper, we studied the discovery potential of the heavy Higgs bosons in Two-Higgs-Doublet Models (2HDMs) at a high-energy muon collider. We take the alignment limit so that the interactions between the Higgs bosons and the SM gauge bosons are of the universal gauge interactions.
We explored the pair production of non-SM Higgs bosons, and single non-SM Higgs production in association with a pair of heavy fermions from both $\mm$-annihilation and the VBF mechanism, as well as radiative return production of a single non-SM Higgs boson directly coupled to $\mm$. We considered the heavy Higgs boson decays to heavy fermions such as the
$t$-quark, $b$-quark, and a $\tau$ lepton.
With appropriate cuts on the invariant mass, transverse momenta, and angular separation between heavy fermions, the dominant SM backgrounds can be effectively suppressed to a negligible level.

We found that the pair production of the heavy Higgs bosons is the dominant mechanism for $\sqrt{s} > 2m_\Phi$, while the single non-SM Higgs production associated with a pair of heavy fermions  could be important for heavier masses, and in regions of $\tan\beta$ with Yukawa coupling enhancement. We also compared the annihilation production versus the VBF production, and found that VBF processes could be dominating at large center of mass energy and low scalar masses.  Radiative return for the single Higgs boson production, in particular, could be important in the large $\tan\beta$ region of Type-L, extending the mass coverage to the kinematic limit $\sqrt{s} \sim m_\Phi$.
With the pair production channels via annihilation, 95\% C.L. exclusion reaches in the Higgs mass up to the production mass threshold of $\sqrt{s}/2$ are possible when channels with different final states are combined.  Including single production modes can extend the reach further.
We reiterate that the discovery coverage at a muon collider would be quite complementary to that at future hadron colliders, where the signal-to-background ratio is low and the signal identification depends heavily on the final states from the heavy Higgs decays and on the sophisticated multiple variable analyses~\cite{Kling:2018xud,Li:2020hao,Hajer:2015gka,Craig:2016ygr}.

We also assessed the discrimination power of a muon collider on different types of 2HDMs.  With the combination of both the production mechanisms and decay patterns,  we found that while it is challenging to distinguish different types of 2HDMs at the  low $\tan\beta$ region, the intermediate and large $\tan\beta$ values offer great discrimination power to separate Type-I and Type-L from Type-II/F.  To further identify either Type-II or Type-F, we need to study the subdominant channels with $\tau$ final states, which could be sizable in the signal rate in Type-II.

Our analyses were performed on the four general types of 2HDMs, in which a  $\mathbb{Z}_2$ symmetry is imposed such that one fermion species only couples to one Higgs doublet to avoid tree-level FCNC.  There exist scenarios in the literature~\cite{Egana-Ugrinovic:2018znw,Egana-Ugrinovic:2019dqu,Penuelas:2017ikk,Botella:2018gzy,Rodejohann:2019izm}  in which flavor alignment or other mechanisms are adopted to suppress the dangerous FCNC effects. As a result, some of the Yukawa couplings no longer need to be proportional to the corresponding quark masses   and the Higgs decay patterns to heavy quarks can be significantly altered.  The general approach outlined in our analyses, however, still applies.  Different SM backgrounds and  judicious cuts need to be reconsidered for the final states involving the first two generation quarks and leptons.

\begin{acknowledgments}

%%%%%%%%%%%%%%%%%%%%%%%%%%%%%%%%%%%%%%%%%%%%%%%%%%%%%%%%%%%%%%%%%%%%%%%%%%%%%%%%%%%%%%
We would like to thank Zhen Liu,
Yang Ma, Patrick Meade and Keping Xie  for discussions. TH is supported in part by the U.S.~Department of Energy under grant No.~DE-FG02-95ER40896 and by the PITT PACC. SL and SS is supported  by the Department of Energy under Grant No.~DE-FG02-13ER41976/DE-SC0009913.
WS were supported by  the Australian Research Council (ARC) Centre of Excellence for Dark Matter Particle Physics (CE200100008). YW is supported by the Natural Sciences and Engineering Research Council of Canada (NSERC). Some of the computing for this project was performed at the High Performance Computing Center at Oklahoma State University supported in part through the National Science Foundation grant OAC-1531128.
\end{acknowledgments}

\bibliographystyle{apsrev4-1}
\bibliography{references}

\end{document}